# Processes in current sheets responsible for fast energy conversion in the magnetospheric collisionless plasma

A.P. Kropotkin

*Skobeltsyn Institute of Nuclear Physics, Moscow State University, 119992 Moscow, Russia*

Phone: +7(495)9393833

Fax: +7(495)9393553

Email: apkrop@dec1.sinp.msu.ru

**Abstract**. Dynamics of the magnetospheric plasma configuration intrinsically features intermittent slow and fast phases. The fast transition is a nonlinear process – loss of equilibrium which ends up the slow quasi-static evolution. The process is analyzed as a dynamical bifurcation. It appears when marginal stability state is reached in the course of that evolution, either for tearing mode or for ballooning mode disturbances. The resulting force imbalance leads to spontaneous formation of nonlinear kinetic thin current structures. Those are either a pair of slow collisionless shocks or a specific anisotropic thin current sheet embedded in a thicker plasma sheet structure. Both are the sites of intense energy conversion, and they implement fast magnetic reconnection in the magnetospheric collisionless plasma.

*Magnetosphere, plasma, nonlinear systems, bifurcation, current sheet, magnetic reconnection, kinetic simulation*

## 1. Introduction

In the decades since the discovery by N. Ness of the geomagnetic tail, the existence of current sheets (CS) in planetary magnetospheres has become common knowledge. Numerous experimental studies of current structures have been supplemented by the development of theoretical models of these structures, of their internal dynamics and their role in the global magnetospheric dynamics. The subject, however, remains highly relevant. Because of the extreme complexity of the physical processes being also detected during the newest experimental programs (CLUSTER, THEMIS, and others), we are still far from a comprehensive theory.

It seems nevertheless useful to draw some preliminary results. It is interesting to bring in some internally consistent picture the results of research carried out by different authors for a quite long period of time.

The general scheme of such an approach should be based on the fact that the magnetosphere–ionosphere system, which exists in a supersonic and superalfvénic solar wind flow, is an open nonlinear dissipative dynamical system (Nicolis and Prigogine 1977, Haken 1983). In such a system naturally arise almost stationary nonlinear structures, which besides are quasi-one-dimensional: their transverse scale being specified by the kinetic properties of the medium, is very small compared to the global scale of the system.

This type of structures has long been known; they play in particular, the most important role in space gas dynamics, see e.g., (Somov 2006); those are shock waves. In the presence of magnetic field, a MHD shock is a CS. It is well known that the bow shock exists in the magnetospheric system as well. However, the fact that this system is formed by interaction of the solar wind carrying the interplanetary magnetic field, with an obstacle in the form of the planet's intrinsic magnetic field, leads to existence of a number of yet other CS separating the different domains of the magnetic field. Those are the magnetopause and the CS in the geomagnetic tail. Their structure and dynamics will be the focus of this review. They are closely connected with the global dynamics of the system.

It is known that while the main components of the magnetosphere–ionosphere system are conserved in time, their location, size, and other properties suffer strong variations. These variations manifest themselves as geomagnetic and auroral activity (Schindler 2006). The corresponding global dynamics of the system is due both to strong non-stationarity of the solar wind and IMF vector, and to the transformation processes, inherent in a nonlinear dissipative dynamical system (Nicolis and Prigogine 1977, Haken 1983, Kropotkin and Sitnov 1997).

The most important features of these dynamics are manifested in characteristic energy transformations. Among them the substorm cycle is of particular importance: relatively slow storage of the magnetic field energy in the geomagnetic tail (loading), occurring on the substorm growth phase, is followed by the fast transformation of that energy into the energy of plasma flows and heat (unloading) which occurs on substorm expansion and recovery phases. These energy transformations are closely related to the processes in the CS, both at the magnetopause and in the geomagnetic tail, which are generally referred to as the processes of magnetic reconnection. We try to show, however, that this is a whole set of different by physical nature, but interrelated processes.

Models of stationary reconnection which already appeared in the 60-ies of XX century (by Sweet, Parker, Petschek, and others, see, e.g. (Priest and Forbes 2000, Schindler 2006)) proposed certain directions for future research. They showed that the solution of the problem must be sought in the analysis of the processes occurring in a relatively thin CS, so that its small transverse scale could allow for a significant diffusion of the magnetic field, resulting in the violation of its frozen-in property. However, rough assumptions were inevitably made which were necessary to link the processes occurring in the diffusion region with those in the convection region, providing plasma outflow from the zone of magnetic flux reconnection.

In addition, long ago it became clear that relatively slow, quasi-static reconnection, which can be reproduced by global stationary models, may not be an adequate explanation for the processes occurring in the geomagnetic tail during the expansion phase of a substorm.

From the theoretical point of view, the controversy arose in the form of convection crisis, see e.g. (Schindler 2006, p. 380). Quasi-stationary convection would lead to such an accumulation of plasma in the near-Earth portion of the plasma layer, which would violate the pressure balance in this area, i.e. would give rise to a withdrawal of the system from the zone of slow, quasi-static reconfigurations in its phase space, which violates the initial assumption of quasi-stationarity.

From the standpoint of observation, it became clearly obvious that indeed a relatively slow restructuring of configuration is typical for the substorm growth phase only, while on the expansion phase it is replaced with fast changes on medium spatial scales occurring with plasma and fields, see e.g. (Baker et al 1999).

Issues that had thus arisen, may be divided into two different groups.

First, it was necessary to understand in principle how to construct a mathematical model which would involve two different time scales, slow and fast, for the behavior of dynamical systems under a slow weak external action, specifying its quasi-static evolution. This action, driving the system through a continuous series of equilibrium configurations, should move it into such a zone in its phase space, where a fast nonlinear process of equilibrium loss takes place.

Second, it was necessary to identify those specific physical processes that are absent on the slow, quasi-static phase, but start to work on the phase of fast evolution.

In what follows, we will focus on presenting the current status of each of these issues in sequence.

## 2. Loss of equilibrium – the dynamical bifurcation

### 2.1. Introductory notes

One of the key issues that remain controversial at present, is as follows: in what time sequence the events occur in the near-Earth and in the remote parts of the geomagnetic tail during the phase of fast evolution, i.e. during the substorm "explosion"; from which part of the tail this fast process is initially induced (Liou et al 2000, Lui et al 2008, Mende et al 2007)? It is believed that the two possibilities correspond to two different physical mechanisms of the initial, triggering process – the equilibrium loss in the tail. In the first case, this mechanism is based on perturbations of the ballooning mode type, and in the second case – of the tearing mode type. Current experimental studies aimed at clarifying this issue are concentrated around the THEMIS project.

Along with this unresolved problem, the most important question, which is closely associated with it, concerns the nature of the equilibrium loss itself as a "fast" process in the magnetosphere–ionosphere nonlinear dynamical system. Such fast processes (unloading), as already mentioned, are alternated with the processes of slow, quasi-static evolution (loading).

The slow variation of equilibrium in the geomagnetic tail during the substorm growth phase is not itself able to bring the system in a highly unstable state, which could manifest itself in a sharp substorm "explosion", i.e. the breakup. The key to solving the problem is the nonlinear dynamics of the system. We propose the particular approach to the problem.

When in the course of quasi-static evolution of the system on the growth phase, it approaches the marginal stability (i.e. a state with an infinitely small growth rate of linear instability that corresponds, in the linear approximation, to infinitely slow, and not a fast evolution), and then leaves that marginal stability point behind, it gets into the zone of so-called catastrophe of equilibrium states (Arnold 2004, Poston and Stewart 1996). Here the particular type of the marginal mode – the ballooning or the tearing mode, and the specific nature of the subsequent rapid nonlinear process – the dynamical path in the zone of that catastrophe, i.e. the course of temporal evolution of the system parameters may be different, as well as the corresponding short characteristic time scales.

## 2.2. Nonlinear ballooning disturbance

The slow, quasi-static variation of equilibrium on the substorm growth phase leads to the formation of a small, in spatial extent, transition region between the quasi-dipole and stretched into tail field lines, at the near-Earth edge of the plasma sheet in the geomagnetic tail. For experimental evidence see, for example, (Kivelson and Russell 1995), and for theoretical considerations – (Kropotkin and Lui 1995). Just in this transition region the ballooning perturbation can develop. Such a configuration instability, developing at the inner edge of the plasma sheet, generates a rarefaction wave (Kropotkin 1972) which propagates into the central region of the tail. There the magnetic disturbance occurs to be sufficient for later appearance of a neutral line, and then – of the magnetic reconnection (Lui 1996).

### 2.2.1. Substorm detonation

With respect to the ballooning perturbation at the near-Earth edge of the plasma sheet in the geomagnetic tail, a model was elaborated of so-called "substorm detonation" (Hurricane et al 1997, Fong et al 1999, Hurricane et al 1999, Wilson and Cowley 2004). Perturbations are considered with the transverse scale much smaller than the longitudinal one. Then the longitudinal structure of the perturbation along the field line is specified as satisfying the Alfvén resonance condition. The eigenvalues of the frequency and their imaginary parts, i.e. the growth rates, are calculated, and thus the possibility of instability in the linear approximation is determined. Ballooning mode equation can be solved for individual field lines, one after another. But this is a linear approximation. Actually, a separate field line cannot move without dragging the neighboring field lines. And the equation of "detonation" is formulated so that it defines this particular nonlinear relationship between the field lines; thus the structure of the perturbation transverse to the field lines is determined, as well as the system evolution over time.

Later theory and simulation for the ballooning mode were developed in more detail. The MHD linear theory was constructed and simulation was carried out for a realistic model of the initial equilibrium involving a $\beta \sim 1$ region (Cheng and Zaharia 2004). Low-frequency kinetic modes at $\beta \sim 1$ and their interrelationship were investigated (Horton et al 2001, Crabtree et al 2003). Finally, the nonlinear MHD simulation was carried out, and the successive phases of perturbations development were revealed (Zhu et al 2007, Zhu et al 2008).

The nonlinear equation of the "substorm detonation" predicts a singularity – an infinite growth of the perturbation amplitude in a finite time, i.e. the equilibrium loss looks like an explosion instability. However, contrary to the idea of "substorm detonation", in the direct MHD simulation a singularity at a finite time interval is not found. It is found that the perturbation remains finite throughout all the nonlinear stage, even with some reduction in the growth rate as compared with the linear stage.

We have suggested an explanation for this failure with the idea of "substorm detonation" (Kropotkin 2012). Following that paper, we present below an analysis of existing differences between the real system and the model leading to "substorm detonation". Those differences prevent the system evolution over the "substorm detonation" path.

Afterwards we return to the aforementioned alternative, that of dynamical bifurcation.

*2.2.2. Problem of small scale over y: phase mixing during reconfigurations*

According to the theory of substorm detonation, the spatial structure of perturbations in a direction transverse to the field lines, evolves to a very narrow finger-like form. The instability begins at a certain local area. At low amplitude two different nonlinear effects are significant. One of these nonlinearities (the explosive nonlinearity) leads to an explosive rise of the mode and formation of finger-like structures, narrow in the direction transverse to the magnetic field. The other nonlinearity (the quasi-linear effect) leads to flattening of the profile and to propagation of the disturbance into the region of linear stability. But the domain of linear stability is in fact metastable, due to the explosion instability. As the "fingers" extend, they lead to destabilization of the metastable region, i.e. "detonation" takes place in the plasma.

It may be seen, that the quasi-linear effect is strongly dependent on the degree of coherence of disturbances on the neighboring field lines: the profile flattens out only when this coherence exists. But this situation can easily get violated. For small-scale perturbations of the transverse uniformity, phasing is easily violated for disturbances in the neighboring regions separated by a distance $\approx k_y^{-1} = \mathrm{O}(n^{-1})$, where $n$ is the azimuthal wave number. Really, there is an additional factor: it is necessary to take into account that the instability arises on the background of continuing evolution of the configuration as a whole, which is characterized by velocities of transverse displacement $v_{conv} \ll V_A$.

What is required for the condition of the longitudinal Alfvénic resonance along the magnetic flux tube to be maintained in spite of its deformation with its length changing, which occurs during the large-scale evolution of the configuration? This to be valid, the deformation must be very slow: the time scale of the flux tube displacement (occurring at a rate $v_{conv}$) in the transverse direction by a value on the order of magnitude of its transverse scale, i.e. the transverse wavelength $2\pi / k_y$, must be large as compared to the characteristic time $\tau_A = L/V_A$, where $L$ is the longitudinal length scale of the field line:

$$\frac{2\pi}{k_y v_{conv}} \gg \frac{L}{V_A}.$$

And if the opposite inequality is valid,

$$\frac{v_{conv}}{2\pi V_A} k_y L \gg 1 \qquad (2.1)$$

the resonance condition on a given field line is violated. And thus the phasing of disturbances on the neighboring field lines is also violated. We see that a relatively fast large-scale restructuring that satisfies the condition (2.1), can break down the growth mechanism for small-scale perturbations with $k_y L \gg 1$.

It can be concluded that the nonlinear instability of small-scale perturbations is suppressed in the non-stationary system of the geomagnetic tail during its evolution taking place on the growth phase of a substorm.

*2.2.3. Thin transition region: a surface wave. The discrete spectrum of perturbations on intermediate scales instead of a continuum over x*

Small-scale perturbations, which were discussed above and for which the theory of linear and nonlinear instability was developed in the works cited above, have a continuous spectrum; it can be said that they form a continuum of oscillations of individual field lines.

The situation looks quite different when we turn to mesoscale disturbances with transverse scale of the order of several $R_E$. At the end of the growth phase, between the quasi-dipole and the tailward-stretched field line domains of the equilibrium configuration on the night side, as already mentioned, there is a relatively sharp boundary, its thickness is $\delta \sim 1 R_E$. Then in this area (where most expected is the nonlinear instability of ballooning modes, because of high radial pressure gradient appearing in that region) for mesoscale perturbation, instead of the continuum of field line resonances, we have a discrete set of oscillation eigenmodes. In the radial direction these are evanescent waves, their amplitude decreases with distance from the boundary. That is, they all have the form of a surface wave in this direction. And in the azimuthal direction, there is a discrete set of modes with different scales $\sim 2\pi / k_y$ (large ones, on the order of several $R_E$). And here, for the surface modes, we have correspondingly $\kappa \sim k_y$.

*2.2.4. An alternative: the dynamical bifurcation*

With regard to such a situation, the theory of so called "dynamical bifurcation" (Neishtadt, 1987,1988, Neishtadt, 2007, Berglund and Kunz, 1999, Berglund, 1999) may be proposed to apply. The basics of this approach, but applied to the tearing perturbations in the CS, were first presented in (Kropotkin, Trubachev and Schindler, 2002a, 2002b). The approach can also be used here since in the layout presented above, from the mathematical point of view, we have a similar situation. This approach is based on the notion that (1) there is a discrete spectrum of linear perturbations; (2) during the slow, quasi-static variation beginning from the state when all the eigenfunctions of that spectrum are stable, the system evolves to the state when one of them becomes marginal, and then unstable. In the situation of the near-Earth plasma sheet, this type of evolution path for ballooning perturbations follows from the continuous change occurring in the course of slow, quasi-static variation, in relationship between the *increasing destabilizing effect* and the stabilizing effect which remains more or less the same. The former is due to the increasing inward plasma pressure gradient, and the latter – to the conducting ionospheric "roots" of the field lines, see, e.g., (Schindler 2006).

The conducting ionospheric "roots" of the field lines form a *constraint* forbidding transverse mutual motions of field tubes. Note however that in contrast to simple mechanical analogs, such motions are not completely forbidden in a distributed system, i.e. in a medium like plasma, in the sense that *fluctuations* are possible. It is well known that thermal fluctuation intensity is extremely high for some specific modes of disturbance, namely for *resonances,* i.e. for intrinsic plasma waves $\omega = \omega(\mathbf{k})$. If the medium is *active*, the corresponding collective motion is no more a fluctuation with some limited spectral power, but is, in the linear approximation, an exponentially growing disturbance – the mode is unstable. Under conditions (1) and (2), see above, as the system quasi-statically evolves, it may at some moment become *marginally active*, i.e. a marginally stable mode appears.

The next Section 2.3 is devoted to this general approach, and to results obtained therein with respect to the equilibrium loss. To the nonlinear tearing perturbation we return in Sec. 2.4.

## 2.3. Model of dynamical bifurcation for the equilibrium loss in a distributed system with discrete spectrum of perturbations

### 2.3.1. Linear analysis near the marginal stability point

Consider a distributed system with its dynamics depending on the parameter $\Lambda$. The dynamics can be represented with some operator $\hat{D}$ acting in the space $(\mathbf{r},t)$ on the vector field $\mathbf{A}(\mathbf{r},t)$ and additionally depending on time through its dependence on $\Lambda(t)$:

$$\hat{D}\big(\mathbf{A}(r,t),\mathbf{r},t;\Lambda(t)\big) = 0, \qquad (2.2)$$

the problem should be complemented with appropriate boundary conditions.

For each fixed $\Lambda$ value, there are some equilibrium states of the system, in general, both stable and unstable. In general case, the linearized version of the system, which describes a small global deviation $\mathbf{a}(\mathbf{r},t) = \mathbf{a}(\mathbf{r})e^{-pt}$ from the equilibrium state $\mathbf{A}_0(\mathbf{r})$ (i.e. solution of the linearized problem (2.2) for some fixed $\Lambda$ value), leads to the characteristic equation $P(p;\Lambda) = 0$, where $P$ is a polynomial on $p$. The degree of the polynomial is determined by the number of degrees of freedom, this number can be large and even infinite. The roots $p_m(\Lambda)$ of the characteristic equation form the eigenvalue spectrum of the system; together with the given boundary conditions they also determine the spatial structure of the corresponding (orthonormal) eigenfunctions $f_m(\mathbf{r};\Lambda)$. The general solution can be written down as the linear combination $\mathbf{a}(\mathbf{r},t;\Lambda) = \sum \mathbf{a}_m(0;\Lambda) f_m(\mathbf{r};\Lambda) e^{-p_m(\Lambda)t}$, where the values $\mathbf{a}_m(0)$ are determined by initial conditions.

### 2.3.2. The dynamical evolution of the marginal mode

In the real situation, with a slowly varying parameter $\Lambda$, for possible fluctuations having a structure distinct from that of the marginal mode, a slow variation of $\Lambda$ appears to be insignificant: the fluctuation disappears completely earlier than $\Lambda$ gets changed significantly.

A quite different behavior is demonstrated by a fluctuation with its structure close to that of the marginal mode, in which case $p_m(\Lambda) = 0$. Such a structure may be determined by solving the corresponding variational problem for the given operator $\hat{D}(\mathbf{A}(r,t),\mathbf{r},t;\Lambda)$, see, e.g. (Schindler et al 1973). Let the marginal mode appear at $\Lambda = \Lambda_*$, with $\mathbf{f}_{\mu,*}(\mathbf{r})$ being the orthonormal eigenfunctions corresponding to $\Lambda = \Lambda_*$. For every $t$, let us present the perturbation $\mathbf{u}(\mathbf{r},t)$ by means of an expansion over that particular basis. The coefficients of expansion – the amplitudes of the linear problem – form the vector

$$\xi = (\xi_1, \ldots \xi_\mu, \ldots),$$

with $\xi_1$ being the amplitude of the marginal mode. Introduce the notation $\xi_1 \equiv a$. For the marginal "slow" mode the following nonlinear dynamical equation may be obtained:

$$\frac{da}{dt} = -\gamma_1(\Lambda(t))a + B_{11}^{(1)}a^2 + C_{111}^{(1)}a^3 + \sum_\mu B_{1\mu}^{(1)}a\xi_\mu$$

$$= -\gamma_1(\Lambda(t))a + \left(B_{11}^{(1)} + \sum_\mu \frac{\beta_{\mu 1}B_{1\mu}^{(1)}}{\gamma_\mu}\right)a^2 + \left(C_{111}^{(1)} + \sum_\mu \frac{B_{11}^{(\mu)}B_{1\mu}^{(1)}}{\gamma_\mu}\right)a^3.$$

The coefficient of the linear term may be presented in the form

$$-\gamma_1(\Lambda(t)) = \Gamma(t-t_1).$$

Then the linear equation describing the dynamics of the marginal mode, may be presented in the following brief form:

$$\frac{da}{dt} = \Gamma(t-t_1)a + \kappa a^2 + \mu a^3. \qquad (2.3)$$

### 2.3.3. Catastrophe of equilibrium

Consider a set of stationary solutions $da/dt = 0$ in the vicinity of the marginal point $\Lambda = \Lambda_*$. What is the full set of such equilibriums under the condition $\Lambda - \Lambda_* \to 0$ ? Setting the left-hand side of (2.3) equal to zero, we see that for each fixed $\Lambda$ in the vicinity of the marginal point, there exists at least one, namely, the trivial equilibrium $a = 0$. However, variation of $\Lambda$ in this neighborhood may be followed by a change of the total number of possible states of equilibrium. Thus we have the simplest, one-dimensional case of singularity on the set of smooth mappings, generating structurally stable set of equilibriums, i.e. a catastrophe of equilibriums (Arnold 2004, Poston and Stewart 1996). Additional equilibrium states existing alongside the trivial state $a = 0$ and different from it, have a spatial structure similar to the structure of the marginal mode, as far as the corresponding values $a \neq 0$ remain sufficiently small. It is necessary to analyze the set of such equilibriums that arise in the generic situation, but at $\Lambda$ values close to $\Lambda_*$.

The specific type of the catastrophe may be inferred from the particular type of bifurcation at the point $\Lambda = \Lambda_*$. The transformation of the equilibrium state $a = 0$ from stable into unstable occurs at bifurcation of the transcritical (stability exchange) type. That bifurcation is not possible in the case of the fold-type catastrophe, so that we have here the only alternative, namely, the cusp catastrophe. This provides valuable fully general information about all the equilibriums existing in the vicinity of the marginal point. The general statement is that the set of equilibriums is given by the cubic form $\gamma_1 a + \kappa a^2 + \mu a^3 = 0$. Without going into detail, we present here the final results of the analysis. They are illustrated in Fig. 2.1. The set of equilibriums $da/dt = 0$ on the plane $(\gamma_1, a)$ is a combination of a straight line $a = 0$ and a parabola $\gamma_1 + \kappa a + \mu a^2 = 0$. With the increase of $\gamma_1$ beginning with negative values, first, at the point (1), a pair of new parabolic branches appears, one stable and another unstable, in addition to the existing equilibrium state $a = 0$, so that it becomes metastable. Further, at the point (2), $\gamma_1 = 0$, the state $a = 0$ becomes unstable, but another stable branch appears in addition. Just here the stability exchange bifurcation occurs which, as already mentioned, provides the conversion of equilibrium $a = 0$ from stable to unstable.

Thus the result of the constructed approximate theory, which is expressed in equation (2.3) for the dynamical nonlinear evolution, has a more general and rigorous mathematical justification in the following sense. If the evolution is a sequence of quasi-static changes, alternating with fast "jumps" of the system from one equilibrium branch to another, these quasi-static changes and these branches near the marginal point are given in the generic situation by a simple catastrophe of the cusp type, as shown in Fig. 2.1.

*2.3.4. Dynamical bifurcation in a one-dimensional system*

But does nonlinear dynamics really have the character of a sequence of quasi-static changes alternating with fast "jumps"? For the one-dimensional problem to which we have reduced our issue, a positive answer to this question is given by the mathematical theory of dynamical bifurcations, elaborated in the last decades (Neishtadt 1987,1988, Neishtadt 2007, Berglund and Kunz 1999, Berglund 1999). In our case it is easy to analyze the solutions of Eq. (2.3).

For small $a$, i.e. near the bifurcation at $t - t_0 \to 0$, it appears sufficient to consider the solution of a simplified, truncated equation, with the third order term neglected. Its solution may be written in the form

$$A = -\frac{e^{\tau^2/2}}{\int_\Theta^\tau e^{\eta^2/2} d\eta}$$

using the following notations: $\tau = \sqrt{\Gamma}(t-t_1)$ and $A = (\kappa/\sqrt{\Gamma})a$. For $|\Theta| \gg 1$

$$A \simeq \frac{1}{-\frac{1}{\tau} + \frac{1}{\Theta}e^{-(\tau^2 - \Theta^2)/2}} \qquad (2.4)$$

It is also easy to find the asymptotic formulas for $|\tau^2 - \Theta^2| \leq 1$, for $\tau^2 - \Theta^2 \gg 1$ and for $\tau^2 - \Theta^2 \ll -1$, and to construct the integral curves. In Fig. 2.2 the integral curves are presented already for the full, non-truncated equation; they tend to the equilibrium asymptotes shown by dotted lines. Fig. 2.2 (a) shows the solutions for different initial values $A(0)$, for a fixed coefficient $R$ in the term cubic over $A$, and Fig. 2.2 (b) shows the dependence of the solution form on the value of that parameter, – at a fixed initial condition.

We are interested in the "fast" parts of the solution, with characteristic times much smaller than the time scale of quasi-static evolution $\tau_1$. Normalizing time scale is $\Gamma^{-1/2} = |(d\gamma_1/dt)_*|^{-1/2} \sim \gamma_m^{-1/2}\tau_1^{1/2}$, here $\gamma_m \sim |(d\gamma_1/dt)_*|\tau_1$ is the typical value of the linear growth rate.

The solution of our nonlinear equation may not be characterized by a single time scale, as is the case with the linear theory. However, here it is also possible and useful to highlight some characteristics with the dimension of time. One of these characteristics is the delay time, i.e. the time of "pulling back" of the jump,

$$T = |\Theta|\Gamma^{-1/2}.$$

As we see, however, this feature depends on the value of initial disturbance. For small $A(0) = A_0$, we have $A_0 \simeq \Theta e^{-\Theta^2/2}$ so that $T$ is determined by the formula

$$\frac{\kappa}{\sqrt{\Gamma}}A(0) = T\sqrt{\Gamma}\exp(-\Gamma T^2/2).$$

It is seen from (2.4) that fast, explosion-like variation of $A$ occurs at $\tau \simeq |\Theta|$ if $|\Theta| \gg 1$. The duration of that jump is the second characteristic time, and it is estimated as

$$\Delta t \sim |\Theta|^{-1}\Gamma^{-1/2}.$$

At larger delay times of the jump, i.e. for smaller initial disturbances, we obtain shorter duration of the jump itself. At $\tau \to \Theta$ the solution has the character of an explosion instability: $A \simeq 1/(\Theta - \tau) \to \infty$ at $\tau \to \Theta$. Clearly, if we take into account the third-order term, we get a fast transition on the time scale $\Delta t$ instead of singularity.

Note also that the specific mechanism of nonlinearity appears to be of small importance: the characteristic time scales $T$ and $\Delta t$ depend on the coefficient of the nonlinear term nearly logarithmically only.

## 2.4. Nonlinear tearing disturbance

### 2.4.1. Introductory notes

The slow, quasi-static variation of the equilibrium on the substorm growth phase, along with forming a small in extent transition region between quasi-dipole and stretched into the tail field lines (see Section 2.2) also leads to the CS thinning.
Configurations of the geomagnetic tail may be characterized by the same value of the total magnetic flux $\Phi$ in the lobes of the tail or the same value of the total magnetic energy, but differ in structure – involve "magnetic islands" or not. On a number of models (Baker et al 1999, Kiessling et al 1986, Schröer et al 1994) it was shown that there is a critical value $\Phi = \Phi_{cr}$ such that at $\Phi < \Phi_{cr}$ a unique equilibrium exists while at $\Phi > \Phi_{cr}$ there is a pair of equilibrium states: one stable and another unstable; they differ in the total entropy value. The point $\Phi = \Phi_{cr}$ is a bifurcation point.

One could assume that under quasi-static growth of the control parameter $\Phi$ corresponding to magnetic "loading" of the geomagnetic tail on the growth phase, after leaving behind the bifurcation point, the system smoothly passes onto the stable, possessing higher entropy, branch with magnetic islands. But this does not happen. Such an evolution requires that dissipation occurs in the system, and it is prohibited in the collisionless limit: the magnetic flux frozen-in condition continues to be valid, so that the magnetic topology changes remain forbidden: "non-ideal" degrees of freedom are suppressed. Therefore, the system evolves along the unstable – in the thermodynamical sense – branch.

The magnetic flux frozen-in condition is a *constraint* forbidding motions involving the "non-ideal" degrees of freedom. Again however, like in Subsec. 2.2.4, there exist resonant fluctuations; and if the medium is *active*, the corresponding collective motion is no more a fluctuation with some limited spectral power, but is, in the linear approximation, an exponentially growing disturbance – the mode is unstable.

As the system quasi-statically evolves starting with initial stable equilibrium, it may at some moment become *marginally active*, i.e. a marginally stable mode appears. Such a situation is similar to what we have considered in Subsec. 2.2.4, with regard to the near-Earth plasma sheet. We should just find out which particular mode of disturbance could in this case first appear as the marginal mode implementing the dynamical bifurcation transition from that unstable branch to the new stable one.

As already mentioned, theoretical and experimental studies (Baker et al 1999, Schindler 1998) indicate an intense CS thinning in the course of evolution, during the magnetic "loading" of the tail. So it can be considered an essential consequence of the quasi-static equilibrium adjustment, also see further, Sec. 3.1.

It may be argued that when the CS thickness becomes sufficiently small, the current density becomes high enough for "non-ideal" processes to enter the game, in other words, the "non-ideal" degrees of freedom are no longer suppressed.

These "non-ideal" degrees of freedom are associated with tearing mode disturbances. For a long time it was assumed that such an equilibrium loss results of the tearing instability. Therefore, a large number of papers were devoted to the possibility of such instability in collisionless plasma at $B_n \neq 0$ (see, e.g., (Pellat *et al* 1991)).

But now it may be considered to be established that such an instability as itself, in its linear form, as well as the governing physical mechanisms, are not fundamentally important. That is because the system never exists in a state "prepared" for the linear tearing with a finite growth rate, and thus it never evolves in the linear tearing instability regime. The current filamentation, the formation of neutral lines and the magnetic flux reconnection – all of them occur at once, already in the nonlinear regime of dynamical bifurcation.

However the existence of the system marginal state and of the marginal mode for linear tearing disturbance are of fundamental importance. Only when that state is reached, the dynamical bifurcation occurs. Therefore, the works that have revealed the viability of such a state, retain their significance. A very valuable point is that use can be made here of the variational "energy" principle, without going into detail of the processes.

Analysis of the collisionless two-dimensional structures leads to the variational principle (Schindler et al 1973): for small variations of the vector potential $A$, the quadratic functional of "energy" $W_0$ can be constructed, so that

$$W_0^* = \inf W_0 = 0 \qquad (2.5)$$

is the condition of marginal stability over the tearing mode, and this condition specifies a certain – marginal – mode.

### *2.4.2. Existence of the marginal mode*

Let the continuously varying parameter, for example, the CS thickness $L$, define the equilibriums set in the vicinity of marginal stability. Then we might have steady states, $W_0^* > 0$ at $L > L_0$, unstable ones, $W_0^* < 0$, at $L < L_0$, and the marginal, $W_0^* = 0$, at $L = L_0$. But it is known that at $B_n \neq 0$ and $T_e/T_i \sim 1$, the requirement of the variational principle for instability, $W_0^* \leq 0$, is not feasible due to the stabilizing effect of electron compressibility, see, for example, (Pellat et al 1991). However, the condition $T_e/T_i \ll 1$ considerably reduces that effect (Sitnov et al 1998), and it can be completely suppressed at low $L$ values. In addition, it is quite likely that (2.5) is generally achievable for $K_x L_x \sim 1$ ($L_x$ is the length scale of the equilibrium configuration over $x$, $L_x/L \gg 1$, and the large-scale disturbance is taken in the form $\sim e^{iK_x x}$). So we consider here that the marginal state is either reached at some critical moment, or the system reaches at least a state very close to the marginal.

### *2.4.3. The one-dimensional evolution*

Like in the situation described in Sec. 2.2, when the control parameter, in this case the $L$ scale, is close to the critical value, the structure of a small disturbance, such that it does not undergo fast decay, should be similar to that of the marginal mode. Really, for all other disturbances different from that, we have $W_0 > 0$, so that they disappear in a short time. Therefore, the disturbance can be defined by a single function of time, its amplitude $A(t)$. The nonlinear dynamics of such a disturbance is then described by the theory outlined above for the ballooning disturbances, see Sec. 2.2.

In the simplest case, we have a nonlinear tearing instability of a "laminar" kinetic type. In principle it is possible that despite the presence of other modes (with much smaller spatial and temporal scales), which are generated in the CS getting thin, they do not interact efficiently with the tearing mode near the marginal stability state. Then, having reached this state during the slow, quasi-static evolution involving the CS thinning, the system gets into the zone of dynamical bifurcation – of a fast equilibrium loss with a jump onto a new branch of stable equilibrium, differing by addition of a finite-amplitude disturbance in the marginal mode. A new configuration, that disturbance being added, may already involve neutral lines and magnetic islands.

### *2.4.4. Fast equilibrium loss due to catastrophic increase of the turbulence level*

Now consider the situation when in the slowly thinning CS, high-frequency (HF) plasma oscillations are excited by cross-field current instabilities (CCI), see e.g. (Lui et al 1991), however, the system remains stable relative to low-frequency large-scale disturbances of the tearing mode type. So those disturbances do not grow spontaneously, in an autonomous regime. It has been shown, however (Kropotkin, Trubachev and Schindler 2002a, 2002b) that if the system is very close to the state marginal (but still stable!) over that mode, then an intense disturbance with the spatial structure of the marginal mode can be generated by an "external" effect: in such a state the system has the property of resonance, see (Kropotkin, Trubachev and Schindler 2002a, 2002b) for more detail.

The density of the "external" current $j_{HF}$, which creates the "driving force" in the model equation, e.g., (Galeev 1984),

$$\frac{d^2 A}{dz^2} - \left[ K_x^2 - U(z) \right] A = \frac{4\pi}{c} j_{HF}(z), \tag{2.6}$$

which determines the amplitude of the disturbance vector potential, may be the result of action of the dynamical pressure force in the field of HF waves; such a force arises if the HF turbulence is inhomogeneous over $x$ (Galeev and Sagdeev 1973). Inhomogeneous turbulent HF electric field acting in the $x$ direction, produces a ponderomotive force acting on each electron in this direction. Action on the electron of the ponderomotive force $F_{HF}$ causes it to move in the $y$ direction with an average velocity

$$\overline{v}_y = \frac{c}{eB_{z0}} F_{HF}.$$

The corresponding current density is

$$j_{HF} = en\overline{v}_y = \frac{nc}{B_{z0}} F_{HF}.$$

The force $F_{HF}$ itself can be easily calculated for the quasi-electrostatic turbulence (Galeev and Sagdeev 1973). As for the specific mode, in the case of plasma with $T_i/T_e \gg 1$, which is typical for the CS, we believe that the electron-acoustic (EA) mode is dominant (Akhiezer et al 1974, Kropotkin et al 1999). For that mode we have

$$F_{HF}^{(e)} \simeq \frac{\omega_{pe}^2}{8\pi n \Omega_e^2} \frac{\partial}{\partial x} \overline{|E|^2} \sim \frac{\omega_{pi}^2}{8\pi n \omega^2} \frac{\partial}{\partial x} \overline{|E|^2} \sim F_{HF}^{(i)}.$$

The external current $j_{HF}$ generates the magnetic field disturbance which can be presented by its vector potential $A$ (see (2.6)), and naturally, the induced current $j_{ind}$ also appears. It is easy to see that near the marginal state, the induced current is much stronger than $j_{HF}$: under these conditions, $j_{ind}$ is anomalously large!

For the EA mode we obtain

$$j_{HF} \sim \frac{iK_x}{8\pi} \frac{c}{B_{n0}} \frac{\omega_{pi}^2}{\omega^2} |E|^2 \,;\, A \sim \frac{iK_x}{2B_{n0}} \frac{Ld}{\tilde{\Delta}} \frac{\omega_{pi}^2}{\omega^2} |E|^2$$

on the central plane. Here $\tilde{\Delta}$ is the parameter going to zero at the marginal stability state, it is responsible for the above resonance.

Variations of the induced current $j_{ind}$ affect the growth rates. They generate variations of the magnetic field normal component,

$$\delta B_n \sim \frac{4\pi}{icK_x} j_{ind},$$

and they, in turn, produce variations in the dispersion of EA waves. For given $B_n$ and mean ion velocity in the CS $V_0$, there is some specific frequency $\omega_b = \omega(k_b)$, such that $\omega(k_b)/k_b = V_0$. Due to dispersion, only at frequencies higher than that $\omega_b$, the plasma is unstable: the phase velocities are lower than $V_0$, and the slope of the ion distribution function is positive there. For smaller $B_n$, $\omega_b$ becomes lower, see Fig. 2.3, and this yields additional new unstable modes. For those modes $k$ which have been already unstable, the growth rates are becoming higher at decreasing $\omega(k)$ because the growth rate is proportional to $kV_0 - \omega(k)$. So the turbulence level increases and $P_{HF}$ also increases. With increasing $B_n$, we have the opposite effect. This ensures the closure of the feedback loop.

There is also variation $\delta V_0$ of the ion velocity, which could as well influence the growth rate variation, but this may be ignored, since there is a $\pi/2$ phase shift in relation to $\delta P_{HF}$.

The effectiveness of feedback can be evaluated:

$$\delta \gamma \sim 8\pi \frac{L}{\tilde{\Delta} d} \frac{\omega_{pi}^2}{\omega_{LH}} \frac{|E|^2}{B_{n0}^2}.$$

If the linear growth rate of the HF instability is very low, i.e. the instability is near the threshold, then the current $j_{ind}$ variations lead to a strong dependence of the growth rate on the phase of the large-scale disturbance. For the HF mode, which in the course of quasi-static evolution of the system becomes unstable at the moment $t=0$, its growth rate can be presented in the form

$$\gamma_{HF} = (\Gamma t + \kappa W)/2,$$

where

$$\Gamma \sim \left.\frac{d\gamma}{dV_0}\right|_{V_0 = \omega_b/k_b} \frac{\omega_b}{k_b} \Upsilon^{-1}$$

and

$$\kappa \sim \frac{8\pi}{B_{n0}^2} \frac{L}{\tilde{\Delta} d} \frac{\omega_{pi}^2}{\omega_{LH}}$$

are constants, and $W = |E|^2$ is the correlator of the given mode, which determines the pressure $P_{HF}$ and is proportional to it. We can now write down the evolution equation for the energy of the mode:

$$dW/dt = \Gamma t W + \kappa W^2.$$

This equation has the same form as the (truncated) equation (2.3). Its solutions are of such a character as described in Sec. 2.2: they involve quick, abrupt transitions. The amplitude $A$ of the large-scale disturbance just repeats the behavior of the turbulent modes intensity.

When a mode of the HF turbulence becomes unstable and begins to increase, this increase can be influenced by intrinsic nonlinearities inherent to the particular mode, and first of all, the effect of quasi-linear saturation. It can be shown, however (Kropotkin et al 2002b) that if the system is very close to marginal stability state, that effect becomes insignificant. On the other hand, it can be shown (Kropotkin et al 2002b) that the presented scenario requires the earlier quasi-static evolution of the system to be quite slow; for higher rates of evolution the scenario is not effective, so that evolution of the system under the previous scenario becomes more probable, see Subsec. 2.4.2.

### 2.4.5. Triggered process

If an external trigger operates in the system, the equilibrium loss occurs earlier than the equilibrium would be close enough to the marginal state and the mechanisms discussed in Sec. 2.4.2 and 2.4.3 could start. Then the process takes the form of a "rigidly excited" instability, which at the level of large-scale variations looks alike the linear tearing instability, starting from a finite amplitude level; actually it is a rather complicated process (Kropotkin et al 1996, 1997, Sitnov and Lui 1999). Again it is a combination of a large-scale electromagnetic disturbance of the tearing mode type with small-scale plasma turbulence, which is generated by the CCI, and is modulated coherently with the large-scale disturbance.

However, this modulation has a nature different from the previous case. Both components of the disturbance are involved in a feedback loop, see Fig. 2.4.

The increasing tearing mode disturbance gives rise to an induced electric field capable of accelerating ions and forming additional inhomogeneous, modulated ion flows, which generate modulated turbulence by means of CCI. In the presence of such a spatial asymmetry of the CCI-excited turbulence, which is due to the spatial structure of the tearing mode, there appears an average turbulent force; and that force, in its turn, drives an additional drift flow of electrons

$$v_x^{(n)} = 2\frac{c}{eB_n}\gamma^{(e)}\frac{\tilde{k}_y}{\tilde{\omega}}\frac{W^{(s)}}{n_0},$$

where $\gamma^{(e)}$ is the electronic contribution to the linear damping rate of the turbulent mode, $\tilde{\omega}$ and $\tilde{k}_y$ are the typical average frequency and $y$-component of the wave vector, respectively;

$$W^{(s)} = \frac{1}{8\pi}\frac{\partial}{\partial\omega}\left(\omega\varepsilon(\omega)\right)\bigg|_{\omega=\omega(\mathbf{k})}|E_c|^2$$

is the mode energy. This flow counteracts the stabilizing effect of the electron compressibility acting on tearing disturbances.

Since in this case the process of equilibrium loss is based on the nonlinear interaction of disturbances on the "macro" and "micro" scales, the instability can occur only in the presence of background fluctuating electromagnetic fields with a finite amplitude and/or external electromagnetic disturbances. Slow, quasi-static CS thinning is a necessary prerequisite for the onset of instability. As the CS is thinning, the electric field $E_y$ threshold for the instability decreases, and finally instability occurs when this threshold becomes lower than the amplitude of background fluctuations or external disturbances. This means that if we "do not notice" the small-scale variations, we can describe the situation as a catastrophe of equilibrium (here — of the simplest "fold" type): the quasi-static evolution of the system (CS thinning) leads to an abrupt equilibrium loss. This is illustrated in Fig. 2.5. The effective potential energy $W$ stays always positive for infinitesimal disturbances; but for finite perturbations it turns to be negative, while the threshold amplitude becomes very small at CS thinning. The threshold is determined by the balance condition for the two terms in the equation which defines the $x$-directed flows of magnetized electrons in the CS,

$$\gamma_T \langle \tilde{n}_{1e} \rangle + ikn_0 \left( \langle v_{1d}^{(d)} \rangle + \langle v_x^{(n)} \rangle \right) = 0,$$

where $\langle ... \rangle$ denotes averaging over the magnetic flux tube. It is known (Sitnov et al 1998) that in the absence of modulated plasma turbulence, the flow of magnetized electrons produces a stabilizing effect on the tearing instability. However, it appears that this effect is counteracted by the above-introduced additional flow of drifting electrons, $\langle v_x^{(n)} \rangle$; it is caused by violation of the generalized momentum $P_y$ conservation for electrons. Such a violation is a consequence of the resonant interaction of electrons with turbulence excited by CCI, so that it can be presented as the action of the turbulent force. To estimate the effect, the quasi-linear theory may be used. The obtained estimate of the threshold electric field intensity is

$$E_y \sim E_y^t = \frac{n_0 e v_{Ti}^2 \gamma_T L^2}{c^2 V_0}.$$

For typical parameters of the CS near the inner edge of the plasma sheet, this threshold is estimated as $\sim 8$ mV/m; the corresponding relative magnitude of the magnetic field normal component turns to be about 0.05.

From the quasi-linear theory the balance conditions can be also obtained, which determine transport of the momentum and energy at a later stage, when the disturbance amplitude much exceeds the threshold. Small-scale "fast" disturbances form a quasi-equilibrium background, which determines the transport rate at every moment of the "slow" large-scale evolution. Through this balance, it appears possible to ignore the stabilizing factor of electronic compressibility, so that the standard expression for the growth rate of the ion tearing mode turns out to be applicable (Schindler 1974):

$$\gamma \sim \frac{v_{Ti}}{L} \left( \frac{\rho_i}{L} \right)^{3/2}.$$

## 2.5. Concluding remarks to Sections 2.2 - 2.4

In the preceding paragraphs, an approach was substantiated to the problem of the substorm equilibrium loss in the magnetosphere–ionosphere plasma system, based on the idea of dynamical bifurcation. This idea appears to be suitable because the relative spatial scales of disturbances allow to treat them as having a discrete spectrum, as shown in Sec. 2.2 and Sec. 2.4. Analysis of the discrete spectrum problem leads, for disturbances near the point of marginal stability, to an effective one-dimensional system, and just for that case a well-developed theory of dynamical bifurcation exists. In the dynamical behavior of the system, there are fast phases alternating with phases of slow, quasi-static evolution. We have found out that this behavior arises as a result of two factors: (1) the quasi-static evolution can drive the system into vicinity of the state marginally stable with respect to excitation of a large-scale (global) eigenmode, (2) at that stage of evolution, necessarily comes into play a nonlinearity inherent in this type of disturbance.

It is possible to see here the well-known general features inherent in the behavior of nonlinear dissipative dynamical systems (Nicolis and Prigogine 1977, Haken 1983). The fast nonlinear evolution of the system with time is determined by the dynamics equation of the "control parameter" (2.3); it is the amplitude of the only particular (marginal) spatial mode. It can be said that the system exhibits the low-dimensional behavior. On the other hand, in the dynamical equation for that marginal mode nonlinear terms are significant, i.e. its finite amplitude is taken into account. But it is also necessary to take into account the nonlinear terms containing that finite amplitude of the marginal mode as a multiplier, in the dynamical equations for other, strongly damped modes. Thus the well-known "subordination" of the damped modes is expressed.

We have analyzed here the main features of the phase of fast evolution itself: the relationship between the duration of the jump and the time delay – the jump "pull-back"; the (nearly logarithmic) dependence of these characteristics on the level of the initial disturbance is obtained.

As already mentioned, although there are many theories of linear plasma instability in the geomagnetic tail, all of them have little to do with the actual process of equilibrium loss. The original concept is internally controversial itself: if the CS equilibrium is unstable, how can the geomagnetic tail remain quiescent, almost unchanged for a long time before the substorm onset? Conversely, if the CS remains stable throughout this period and only approaches the marginal stability in the course of quasi-static evolution, how can it suddenly become very unstable? The clue to the problem lies in the nonlinear nature of the system behavior. When the quasi-static evolution drives it close to the point of marginal stability, and then passes this point, there occurs a catastrophe of equilibrium, and the subsequent jump-like disturbance is described by nonlinear dynamics.

Disturbance in the marginal mode has a very special character. It corresponds to the neutral equilibrium: in the first approximation, such a disturbance having an arbitrary amplitude, does not change the energy of the configuration. Therefore, it can get formed in an arbitrarily short time, unlike the truly unstable modes, which in the case of the tearing modes are the negative energy disturbances, so that their growth rate is determined by the possible rate of energy dissipation in the CS.

Why in the nonlinear instability, features must be combined of the tearing instability and of the cross-current HF instabilities, why the interaction of these modes is really of fundamental importance? On the one hand, the CCI actually generates high-frequency wave turbulence in the thinning CS, with a high current density. On the other hand, the quasi-static evolution is slow, so there is often enough time for a large triggering fluctuation to occur and the rigid excitation (Subsec. 2.4.5) to take place, giving start to coherent modulation of the generated HF waves. In the absence of such a trigger, as it has been shown, the slow previous evolution contributes to appearance of a combined instability according to the soft excitation scenario (Subsec. 2.4.4). We note, however, that if the previous evolution is not too slow, the presence and influence of HF turbulence may prove to be irrelevant, and then the "laminar" scenario (Subsec. 2.4.3) works.

Our analysis allows to take a fresh look at the long-debated issue of "forced" and "spontaneous" magnetic reconnection.

"Spontaneous" reconnection as the tearing instability of the CS is possible only in situation of "immediately prepared" unstable equilibrium. Realistic scenario of slow, quasi-static evolution preceding the "explosion" phase, is, as already pointed out, the opposite of this hypothetical situation. There is quasi-static thinning of the CS, and this can be interpreted as *"forced"* restructuring of the system. But when the critically small CS thickness is reached, a *non-linear dynamical bifurcation* takes place – a fast process, bringing the system into a new state (with new neutral lines and one or more magnetic islands in the case of tearing disturbance). To describe this process, *nonlinear* in principle as it is, the *linear* theory of tearing instability i.e. of *"spontaneous" reconnection*, is still to some extent feasible, – as a separate "block" of such a description. This is first because the spatial structure of the disturbance corresponds to the *marginal mode* of such instability. Second, the term in the equation describing the dynamic bifurcation, linear over the disturbance amplitude, has the same form as the corresponding term in the equation for linear instability marginal mode. In this case, however, the coefficient of the linear term, i.e. the instability growth rate, is no longer constant, but is a slow function of time, passing through zero from negative to positive values.

Further evolution of the newly-formed quasi-equilibrium configuration with a new neutral line, occurs already on a relatively large time scale and can be described as a forced stationary magnetic reconnection: the reconnection rate is not determined by processes in the vicinity of the neutral line but by external conditions – by the processes of large-scale restructuring of the tail configuration (Kuznetsova et al 2001).

A question might be asked: for the linear tearing instability at $B_n \neq 0$, how do the physical mechanisms operate and how does this relate to the mechanisms operating at reconnection – a large-scale restructuring of the tail configuration (Kuznetsova et al 2001)? The answer would be: the latter has nothing to do with the former. Moreover, as it has been pointed out in Subsec. 2.4.1, the nature of those tearing instability physical mechanisms themselves is not important in this context. The reason is that the system never exists in a state "prepared" for the linear tearing instability with a finite growth rate, and thus never evolves in the linear tearing instability regime. Current filamentation, formation of neutral lines and the initiation of magnetic flux reconnection occur in the nonlinear regime of dynamical bifurcation. And the numerical simulation and the interpretation (Kuznetsova et al 2001) relate to the later stage of that dynamical bifurcation process, where the neutral line has been already formed. Thus, speaking of the equilibrium loss – dynamical bifurcation, we have its physical picture for that stage namely.

However, the existence of the marginal state of the system and the marginal mode of the linear tearing disturbance are crucial, as we have explained above. Only when that state has been reached and then passed, the dynamical bifurcation occurs.

# 3. Features of the specific dynamics of the geomagnetic tail

When considering large-scale, global processes of the magnetosphere-ionosphere dynamics, quasi-stationary and transient processes are distinguished. It is believed that the passage of the magnetosphere-ionosphere system from one steady state to another occurs over the Alfvénic time scale. This scale $\tau$ is estimated as the time of the Alfvén wave propagation, with its speed ~ 1000 km/s in the magnetosphere, over a typical spatial scale of the magnetosphere, of the order of $10^5$ km, and this yields $\tau$ ~ 100 s. Accordingly, at an intermediate scale of a few $R_E$, the transient process should take still much shorter time, $\tau$ ~ 10 s.

However, due to the specific structure of the magnetospheric configuration, with the stretched geomagnetic tail, and the plasma sheet separating its lobes, to characterize all transient processes with a single time scale appears to be incorrect. Indeed, in the plasma sheet the field normal component $B_n$ is small and, accordingly, the Ampère force $[\mathbf{jB}]/c$ which drives the plasma towards the Earth (or outwards), is small. Therefore, the emergence of such motions takes a relatively long time, and this significantly affects the duration of those transient processes that require the involvement of such motions. Consequently, these processes can be considered separately assuming that the evolution that precedes them, "instantly" drives the magnetoplasma system to a non-equilibrium state, which then relaxes relatively slowly. We emphasize that this only applies to that relaxation, which involves the Earthward or anti-Earthward plasma motion in the plasma sheet. Those motions are the necessary constituent of the magnetic reconnection process in the geomagnetic tail: in slow motions the magnetic flux through the central plane of the CS is frozen in the plasma, so the Earthward and anti-Earthward motions occurring simultaneously, must be accompanied by violation of that magnetic flux conservation, i.e. by magnetic reconnection.

We may conclude that because of strong differences in the duration of these processes having a causal relationship and following one after another, a radical simplification of the theoretical description and simulation is possible. Taking advantage of this opportunity with respect to localized bursts of activity that occur in the plasma sheet of the geomagnetic tail, it is possible to propose a scenario of magnetospheric dynamics as a sequence of separate, relatively simple restructurings – transitions between different localized on the spatial scale $\sim R_E$, quasi-equilibrium states of the system.

## 3.1. Quasi-static configurations and equilibrium loss

The situation outlined above on a qualitative level, in essence, is reflected in the existing mathematical models of the geomagnetic tail. The mathematical descriptions of quasi-two-dimensional magnetic configurations with the CS to be used as models of the geomagnetic tail, appeared almost simultaneously in the works by Syrovatskii (1971) and Schindler (1972). However those are different models. In later years, after the early death of S.I. Syrovatskii, only Schindler's model was further developed in numerous studies carried out by him and his co-authors. But, as we have shown in (Kropotkin and Domrin 2009), Syrovatsky's approach can be used effectively at the appropriate place, in combination with the approach of Schindler.

### 3.1.1. Quasi-static adiabatic restructuring

The studies by Birn and Schindler (see (Birn et al 2006) and references therein) show that the quasi-static evolution of the geomagnetic tail, subject to the condition of adiabaticity – conservation of entropy in the field tube, leads to formation of extremely thin (singular in MHD) current sheets (CS). This analysis is made on the basis of the asymptotic solution of the Grad – Shafranov equation, which was first proposed by Schindler for the situation when, in planar geometry, the spatial scale of the configuration in the $x$ direction, $L_x$, far exceeds the scale $L_z$ in the $z$ direction. It was shown that the quasi-static evolution can lead to the appearance of a new, very small scale – the CS thickness.

In that configuration, the isotropic plasma pressure is a function of the only nonzero component of the vector potential, $A_y = A$: $p = p(A)$. In the asymptotic solution, balance of the plasma pressure and the magnetic field pressure is maintained:

$$p(x,z) + \frac{B^2(x,z)}{8\pi} = \hat{p}(x), \qquad (3.1)$$

where $\hat{p}(x)$ is the pressure on the central plane $z = 0$. When the current is concentrated in a thin CS, the dependence $p = p(A)$ has the form shown in Fig.

3.1: the pressure is nonzero at $A \leq A_1$, and is almost equal to zero everywhere at $A > A_1$. For the field line $A = \text{const}$ an expression is obtained

$$z(x, A) = \int_{A_0(x)}^{A} \frac{dA}{\sqrt{8\pi \left[ \hat{p}(x) - p(A) \right]}},$$

where $A_0(x)$ is the potential $A$ profile on the central plane. Assuming that the magnetic flux through the magnetopause is approximately equal to zero, we can write down an expression for the $z$-coordinate of the magnetopause:

$$a(x) = \int_{A_0(x)}^{A_b} \frac{dA}{\sqrt{8\pi \left[ \hat{p}(x) - p(A) \right]}},$$

where $A_b$ is the potential value at the magnetopause.

As shown in (Birn and Schindler 2002), a family of equilibrium configurations can be specified by fixing some model dependence between $a$ and $\hat{p}$, for example,

$$a(\hat{p}) = a_m + (a_\infty - a_m)\left(|\hat{p} - p_m| / p_m\right)^k.$$

Quasi-static transitions between equilibriums belonging to this family, can be modeled by specifying the variation of parameters, such as $a_m$ parameter. Under the condition of adiabaticity applied to those transitions, so that the pressure $p$ in the flux tube $A = \text{const}$ and its specific volume $V$,

$$V(A) = \int_0^{x_0} \frac{dx}{B_x(x, A)},$$

are related by the adiabatic equation, $pV^\gamma = \text{const}$, we can obtain a sequence of locally thinning CS, and a critical situation can be reached: when the CS thickness tends to zero, the current density goes locally to infinity, see Fig. 3.2.

This theory definitely points to formation of thin CS: the non-zero pressure is concentrated in a narrow region near the central plane. How is the density distributed in such a configuration? Consideration of the quasi-static evolution, based on the arbitrary choice of model dependence $a(\hat{p})$, provides no definite answer to this question. To get a qualitative answer is possible from the following considerations. Strong elongation of the tail means that at a given cross-section $x$, the field lines $A = \text{const}$ outside the CS, have a much greater extent, and accordingly, the specific volume than those belonging to the CS. On the other hand, the latter had, in earlier times, just such a great extent. This is indicated by the existence of magnetospheric convection, directed towards the Earth, in the near-Earth part of the geomagnetic tail. It is convection that on average, and on long time scales, determines evolution of flux tubes in the tail – their contraction, accompanied by an increase in density. Thus, in the real situation the existence of a large pressure maximum near the central plane is due to evolution: it is the result of adiabatic compression accompanied by heating, which occurs in the flux tubes convecting to the Earth, and shrinking at the same time. This effect has long been known, and some time ago it gave a basis for discussion of the so-called "convection crisis" (see, for example, (Schindler 2006)). In a given cross-section $x$, the field lines originally belonging to the tail lobes where $p \ll B_0^2/8\pi$ ($B_0$ is the field outside CS) are carried later into the CS. The gas kinetic pressure $p = p(A)$ which has a plasma, brought inside CS, must be equal to $B_0^2/8\pi$, the magnetic pressure outside CS (this is a consequence of constantly maintained balance of the transverse component of momentum (3.1)). This means that during the convective transport, the pressure in the flux tube gets much larger than it has been in that flux tube when the tube was still outside CS where $\beta = 8\pi p/B_0^2 \ll 1$.

Thus, in agreement with modern observations, the dominating on average, relatively slow and large-scale evolution of the geomagnetic tail always supports the existence of a dense and hot plasma layer in its central part, in the plasma sheet, and of a cold rarefied plasma background – in the tail lobes.

As shown by Birn and Schindler on the theoretical and numerical models, and as we have briefly presented above, over this effect are superimposed sporadic localized effects of the plasma sheet thinning (acting on the intermediate scales small compared with the global magnetospheric scales, and large compared with the CS thickness).

### 3.1.2. Rapid restructuring of configuration with equilibrium loss in the CS

Consideration of the further evolution of the configuration that has formed, with a thin CS, can be simplified by keeping in mind that outside the CS the strong magnetic field approximation holds (Syrovatskii 1971, Somov and Syrovatskii 1976),

$$\beta = 8\pi p / B^2 \ll 1, \; M_A = V/V_A \ll 1. \tag{3.2}$$

Then everywhere outside the CS the magnetic field is potential, and the potential $A$ satisfies the two-dimensional Laplace equation, $\Delta_2 A = 0$. As a boundary condition, it can be adopted that $A|_\Sigma = f_1(s)$ where $s$ is the coordinate measured along the boundary $\Sigma$. Since the CS is thin, it can be assumed that the part of the boundary of the strong field domain, which is on the CS outer edge, coincides with the $z = 0$ axis. A simplest two-dimensional model of that type is shown in Fig. 3.3.

Interestingly, in the approximation (3.2) we may deal not only with the slow quasi-static evolution of the entire system, but also consider such relatively fast changes in which only its part located outside the CS, has enough time to readjust quasi-statically. The point is that formally the restructuring occurs instantaneously, following the changes of boundary conditions. In the Syrovatskii model, the propagation speed of interactions tends to infinity, $V_A \to \infty$, and the potential $A$ at any given time, at each point is determined simply by solution of the corresponding boundary-value problem of the Laplace equation.

However, from a physical point of view, the change itself of the boundary conditions cannot be completely arbitrary. The boundary condition at $z = 0$, i.e. the function $f_1(x)$ can change only slowly. In fact, fast time variation of the potential $A = f_1(x)$ at a given point on the $z = 0$ surface would mean an almost instantaneous appearance of fast MHD flow in the CS itself, which cannot be. We have shown above that as a result of the earlier evolution, the plasma in the CS is relatively hot and dense: if $\rho_1$ is the density outside CS, and $\rho_0$ is the average density in the CS, then we have

$$\rho_1 / \rho_0 \ll 1. \tag{3.3}$$

Now we show that under conditions when there is a thin layer along which all quantities vary only on a large spatial scale, and it separates the regions occupied by low-density cold background plasma, a flow has no time to appear on the time scale of fast changes in the system. And if the other boundary conditions do change rapidly, everywhere outside the CS the configuration continues to be approximately described by the Syrovatskii model with $f_1(x)$ unchanged (on the time scale of these fast changes). As a consequence, in the thin CS the longitudinal equilibrium condition appears to be violated. In that condition,

$$-\frac{\partial}{\partial x}\int_{-\infty}^{\infty} p\, dz + \frac{1}{4\pi} B_n B_0 = 0, \qquad (3.4)$$

the second term remains almost unchanged while the first term changes. A loss of balance occurs.

Outside the CS the magnetic field is adjusting by means of propagating fast perturbations. They are MHD waves traveling with a large but finite velocity, $V_{A1} = B_0 / \sqrt{4\pi\rho_1}$, see, for example, (Domrin and Kropotkin 2007a). However, if we consider changes in the region of finite size $L$, then for not too fast disturbances occurring at such a time scale $T_1 = 2\pi/\omega$ that

$$kL = (\omega/V_{A1})L \ll 1, \qquad (3.5)$$

the magnetic field at every instance in that region can be described by equations of the static field, i.e. that field is quasi-stationary. The field polarization is such that on the surface of the CS (small) strengthening and weakening of the tangential component $\delta B_t$ appear. They cause small contractions and expansions of the CS, i.e. fast changes in the longitudinal distribution of pressure in the CS itself. Really, on the CS thickness $d \ll L$, which is very small compared to the wavelength $2\pi/k = 2\pi V_{A1}/\omega \gg L$, a balance of the transverse momentum component establishes at every instance. Therefore, the plasma pressure $p = p_0 + \delta p$ on the central plane of the CS changes by an amount

$$\delta p \sim \left[(B_0 + \delta B_t)^2 - B_0^2\right]/8\pi \simeq B_0 \delta B_t / 4\pi. \qquad (3.6)$$

Inside the CS, these pressure variations generate longitudinal plasma motions. Estimate the magnitude of these variations. Consider a small disturbance as a sum of harmonic waves propagating with velocity $V_{A1}$,

$$\delta p, \delta B_t \sim e^{i(kx-\omega t)}.$$

For a component with wavenumber $k$, the increment in the first term of (3.4) is $\sim (kd)(\delta p)_k$, while in the second term it is

$$\sim \frac{1}{4\pi}\left[B_n (\delta B_t)_k + B_0 (\delta B_n)_k\right]. \qquad (3.7)$$

In a reference frame moving with the wave at a speed $V_{A1}$, the CS is a stationary one-dimensional flow in a thin flat channel along which all the variables depend on one variable $x$. According to the Euler equation

$$\rho_0 V_{A1} \cdot \delta v(x) = -\delta p(x).$$

In view of (3.6) we obtain

$$\frac{(\delta v)_k}{V_{A1}} = -\frac{B_0}{4\pi \rho_0 V_{A1}^2}(\delta B_t)_k.$$

Since $V_{A1} \sim B_0 / \sqrt{4\pi \rho_1}$, we find

$$\frac{(\delta v)_k}{V_{A1}} = -\frac{\rho_1}{\rho_0}\frac{(\delta B_t)_k}{B_0}.$$

In the harmonic wave, during one half-period the plasma element with its length $\lambda/2 = \pi V_{A1}/\omega$ gets stretched (or compressed) by the amount $(\delta x)_k \sim \pi \cdot (\delta v)_k / \omega$. Under the action of disturbance, the field normal component $B_n$, due to the field frozen-in property, should also vary conserving the magnetic flux through the given plasma element. Accordingly, the stretching (compression) of the plasma element by this amount leads to a relative decrease (increase) of the field by the value

$$\frac{(\delta B_n)_k}{B_n} = \frac{2(\delta x)_k}{\lambda} = \frac{(\delta v)_k}{V_{A1}}.$$

As a result we obtain a simple relationship between the relative changes of the normal and tangential field components and the relative variation of pressure:

$$\frac{(\delta B_n)_k}{B_n} = \frac{\rho_1}{\rho_0}\frac{(\delta B_t)_k}{B_0} = \frac{\rho_1}{\rho_0}\frac{(\delta p)_k}{2 p_0}.$$

We see that for a layer of hot plasma embedded in a uniform very low-density cold plasma background, the relative changes of the magnetic field normal component and of the velocity are smaller than the variations of the field tangential component, the variations of pressure and density; their ratio is of the order of the small factor $\rho_1/\rho_0 \ll 1$.

We now estimate the terms in Eq. (3.7). According to the equilibrium condition (3.4), we get

$$\frac{B_n B_0}{4\pi} \sim p_0 \frac{d}{L_x}, \tag{3.8}$$

where $L_x$ is the longitudinal scale of the equilibrium geomagnetic tail as a whole, i.e. the global magnetospheric length scale.

We emphasize that in our consideration, the natural hierarchy of scale lengths is provided (given that $k = 2\pi / V_{A1} T_1$) by the double inequality

$$L_x \gg V_{A1} T_1 \gg L. \tag{3.9}$$

In view of Eq. (3.8) we obtain

$$\frac{1}{4\pi} B_n (\delta B_t)_k \sim p_0 \frac{d}{L_x} \frac{(\delta B_t)_k}{B_0} \sim p_0 \frac{d}{L_x} \frac{(\delta p)_k}{p_0} \sim \frac{d}{L_x} (\delta p)_k. \tag{3.10}$$

For the second term we find

$$\frac{1}{4\pi} B_0 (\delta B_n)_k \sim p_0 \frac{d}{L_x} \frac{(\delta B_n)_k}{B_0} \sim p_0 \frac{d}{L_x} \frac{\rho_1}{\rho_0} \frac{(\delta p)_k}{p_0} \sim \frac{d}{L_x} \frac{\rho_1}{\rho_0} (\delta p)_k. \tag{3.11}$$

At $kL_x \gg 1$ the increment of the first term in the longitudinal momentum balance equation, $\sim (kd)(\delta p)_k$, is much greater than the increment (3.10), estimated as $\sim (d / L_x)(\delta p)_k$. The increment (3.11) is certainly small, under the condition $\rho_1 / \rho_0 \leq 1$, which is valid because of convective plasma "raking" in the CS. Thus, at $kL_x \gg 1$ the contributions (3.10) and (3.11) provide negligible corrections only.

We now assume that the disturbance amplitude remaining small, is still large enough to satisfy the condition

$$(kL_x) \frac{(\delta p)_k}{p_0} \geq 1.$$

Such a disturbance violates the balance (3.4) to a substantial extent locally, on the scale length of the order of the wavelength $2\pi / k$, and hence on a smaller scale, $L \ll 2\pi / k,$ as well, so the original longitudinal pressure gradient can be substantially suppressed by that disturbance.

So, in a short time $T_1$ the longitudinal pressure gradient in the local region of size $L$ can be perturbed and appears quite different from the equilibrium one, and then it is no more compensated by magnetic tension (which varies only slightly), and the longitudinal balance is thus violated. The time scale $T_1 \sim 2\pi/\omega = 2\pi/kV_{A1}$ corresponds to the wavelength $2\pi/k$ intermediate between the global magnetospheric scale $L_x$ and the scale length $L$, see Eq. (3.9).

Further relaxation, locally – on the scale $L$, is determined by the small but non-zero value of the resultant longitudinal force. Because of its smallness, relaxation occurs on a much larger time scale, $T_2 \gg T_1$. The time scale $T_2$ characteristic of the nonlinear relaxation process, is approximately estimated as follows. The ultimate longitudinal momentum balance is achieved by accelerating ions in the plasma sheet under action of the electric field $E_y = E_0$ that arises,

$$E_0 = -\frac{B_n}{c}V_{A1} = -\frac{B_0 B_n}{c\sqrt{4\pi\rho_1}},  \qquad (3.12)$$

accompanied by their rotation in the magnetic field $B_z = B_n$, after which the average longitudinal momentum of the ion becomes equal to $m_i V_{A1}$ (see, e.g., (Domrin and Kropotkin 2007a,b)). Hence, we obtain an estimate of $T_2$. The momentum balance is given by the condition

$$\rho_0 V_{A1} d \sim \frac{B_n B_0}{4\pi} T_2,$$

where $d$ is the plasma sheet thickness. Consequently,

$$T_2 \sim \frac{\rho_0}{\rho_1} \frac{B_0}{B_n} \frac{d}{V_{A1}} \sim \frac{\rho_0}{\rho_1} \frac{L_x}{V_{A1}},$$

since in the initial equilibrium we have

$$\frac{d}{L_x} \sim \frac{B_n}{B_0}.$$

On the other hand, $T_1 \sim 2\pi/kV_{A1}$, so that by virtue of smallness of the parameters $2\pi/kL_x \ll 1$ (see Eq. (3.8)) and $\rho_1/\rho_0 \ll 1$ (Eq. (3.3)), the inequality $T_2 \sim (\rho_0/\rho_1)(kL_x/2\pi)T_1 \gg T_1$ is indeed correct. And with such a strong difference of the time scales $T_1$ and $T_2$, the simplest model of the process will be the relaxation of one-dimensional layer in which the pressure is uniform in the longitudinal direction, but at the initial instance $t = 0$, an uncompensated force of magnetic tension $B_n B_0/4\pi$ appears. In other words, the model system is *prepared* at the moment $t = 0$ as a one-dimensional system, in which the departure from equilibrium is determined by a nonzero magnetic field component $B_n$.

Thus, the large difference between longitudinal and transverse scales, along with the relatively high plasma density inside the CS, which are manifested during the slow adiabatic restructuring of the configuration accompanied by strong CS thinning, represented in Schindler's theory, allows to introduce an important simplification into analysis. Extremely fast adjustments of the external region (the tail "lobes") can be described by Syrovatskii's model. In the plasma sheet they "instantly" generate local pressure nonuniformities and related violations of the longitudinal force balance.

## 3.2. Relaxation of quasi-one-dimensional thin CS and the magnetic field "annihilation"

Analysis of further relatively slow relaxation can be performed on a one-dimensional model, weakly out-of-balance at the initial time. This process has been investigated in a number of different models as an evolution of "reconnection layer", see e.g. the review (Lin and Lee 1994). A detailed numerical study in the kinetic (hybrid) model by means of a particle-in-cell code has been carried out in the papers (Domrin and Kropotkin 2003, 2004a,b, 2007a,b,c).

Let us present such an approach in a broader context. The problem of formation of thin CS and related relaxation of a magnetoplasma system is in general very complicated. An attempt to cover it with the assistance of numerous studies carried out in the recent decades, was made, in particular, in the book (Priest and Forbes 2000). In the approaches dominating in the literature, it is suggested that the configuration as a whole and its dynamics can be adequately described if the existence in the configuration of one or more singularities is accounted for, those of the type of neutral lines and points, as well as associated separatrice, etc. The processes in such a system are interpreted as magnetic reconnection. However, it is admitted that the configuration itself and its dynamics are very sensitive to the nature of boundary conditions. This introduces much additional uncertainty into the problem.

We believe that this approach unfairly ignores the important fact that may have a dominant significance in the real physical system, and at the same time may simplify the theoretical description. Moreover, it takes it to another plane. We have in mind the formation of thin CS, which is in the MHD models represented as the emergence of one-dimensional singular currents. A well-known example of an MHD shock formation indicates that the formation of CS, in general, is not tied to any singularity features such as, say, neutral lines in the initial configuration. In fact, the clue role belongs to nonlinearity which is inherent in the system and manifests itself, in the course of its dynamical evolution, as the steepening of inhomogeneity fronts. In a real system thin CS can only be generated locally, on the scale lengths (along the CS) being small as compared with the global scale. And the arising relaxation processes (including the "spontaneous annihilation" of magnetic fields) occur also locally - over these intermediate spatial scales. Therefore, they should begin and end – within each mesoscale nonuniformity – in a relatively short time. This allows us to represent the entire global dynamic process as a set of multiple individual localized bursts of relaxation, each of them allowing an approximate one-dimensional description.

Thus, the fundamental feature of this approach is that an essentially one-dimensional character is taken into account of those dynamical processes that determine the fast conversion of the magnetic field energy: the conversion takes place on an extended surface of the CS, on a scale very large compared with its thickness, but small as compared to the characteristic size of the geomagnetic tail as a whole.

Theory and numerical simulations provide an opportunity to describe these dynamical processes as the elements of restructuring taking place in the CS and the surrounding plasma after the moment when, as a result of nonlinear instability (dynamical bifurcation) in the plasma sheet, the loss of balance in the system has occurred. Under condition of smallness of the magnetic field normal component $B_n$, restructuring consists of two different phases: a fast and a relatively slow ones. As discussed above, first the longitudinal momentum balance in the tail quickly – on the Alfvénic time scale – is violated by passing along it of a fast MHD disturbance whose form is determined by the CS presence.

The resulting almost one-dimensional out-of-balance structure with the CS, evolves later, on a significantly larger time scale, by CS splitting into several current structures. They can be regarded as arising spontaneously in the course of the system relaxation when it is derived from the balance on the previous fast phase (just like a shock arises in a gas in the course of supersonic expansion of an instantaneous point explosion products (Landau and Lifshitz 1987)).

In the MHD approximation, one-dimensional dynamics for a similar, but highly idealized situation has long ago been studied, see, e.g., (Heyn et al 1988, Semenov et al 1992, Lin and Lee 1994). The problem reduces to solving the well-known Riemann problem of arbitrary discontinuity decay (Akhiezer et al 1974). The resulting finite amplitude disturbance has the form of self-similar expansion wave propagating from the source CS over the background plasma with the Alfvén velocity, followed by a slow disturbance having the form of a shock wave.

It remains, however, necessary to determine the extent to which this MHD model reproduces the picture of the nonlinear disturbances in a specific situation of the geomagnetic tail, with hot collisionless plasma that forms the CS in its central region, and the cold rarefied plasma background. We must also determine if there are evolution paths of relaxation of the out-of-balance system, other than those appearing in MHD, such in particular, where the CS of specific type arise, with plasma being significantly anisotropic on both sides of the CS – in MHD such solutions are in principle absent. To this end, the kinetic theory is needed.

As shown by numerical simulations (Lin and Lee 1994), the evolution under consideration, in general, may include the Hall current effects and generation of the magnetic field $y$-component, which leads to formation, among others, of structures of the rotational discontinuity type. However, under condition $\rho_1 / \rho_0 \ll 1$, the CS evolution is slow enough, so that the Hall current effects may be neglected (see below, Sec. 3.3.4, and (Domrin and Kropotkin 2007c)).

Since the resulting thin CS may have a thickness of about an ion gyro-radius, the major new features should originate of kinetic effects. It should be borne in mind that all magnetoplasma configurations with thin CS are not limited to discontinuous solutions allowed by the MHD. As shown in (Domrin and Kropotkin 2006), configurations may exist with strongly anisotropic ion distributions on the basis of "forced" kinetic current sheet (FKCS). In (Kropotkin and Domrin 1995, Kropotkin and Domrin 1996, Kropotkin et al 1997, Sitnov et al 2000) the self-consistent kinetic theory of FKCS was first constructed; it incorporated the ideas of the earlier studies of Speiser, Eastwood, Hill, Frankfort and Pellat, Burkhardt et al., Holland and Chen, Pritchett and Coroniti.

Considerable literature is devoted to dynamics of the configuration with CS, as viewed in the kinetic approximation. Many papers are based on two-dimensional and even three-dimensional numerical simulation. Their brief overview is contained, for example, in the articles (Arzner and Scholer 2001) and (Pritchett 2005). However, in most cases, the models with which the dynamics is considered, are restricted to some arbitrary spatial-temporal areas of the system behavior. The proposed boundary and initial conditions are also respectively different, arbitrary and often artificial. As expected, the results are significantly various, and their interpretation and elucidation of their degree of reliability appear very difficult.

Some features of the previous approaches can be more clearly identified with the use of an example of the paper (Arzner and Scholer 2001). In that paper, the study was carried out by means of hybrid simulation for the system, which is a two-dimensional analog of the geomagnetic tail. The process of magnetic reconnection is triggered artificially with the help of an area with anomalous conductivity introduced locally in the CS. A neutral line (NL) arises, and the process occurring in its close vicinity, takes on those traits (role of the Hall current, appearance of the magnetic field $y$-component, etc.) which, in a similar hybrid simulation scheme, had already been obtained for that area in (Kuznetsova et al 1996). These properties were then confirmed in the simulations, carried out by other groups, with other codes (see the final article on the coordinated work, (Kuznetsova et al 2001)), so that we have already reliably established pattern of the magnetic reconnection process in that area.

As stated in Sec. 2.5, such a pattern of the magnetic reconnection process can appear also without the artificially introduced area with anomalous conductivity. In reality, appearance of the NL and the subsequent evolution of the disturbance in its vicinity may be a part of the process of dynamical bifurcation in the thin CS.

Along with these results, in the paper (Arzner and Scholer 2001) results were also obtained relating to the area remote from the NL. A very thin CS arises, located between the NL and the plasmoid receding from it. It is found that acceleration of ions on quasi-adiabatic orbits occurs in that CS. Outside the CS accelerated ions form a fast flow along the field lines, i.e. a significant ion anisotropy appears.

These essential kinetic features that arise during the evolution of the system, were demonstrated by the obtained ion distribution functions. These basically important results that are confirmed also by our work, were obtained however under some cardinal constraints arising from the approach of the authors. Artificially introduced finite conductivity zone generates a single disturbance in the whole system in which all scales are coupled. This circumstance seems to have been amplified by a small number of spatial grid nodes used in the calculations. Therefore, the model failed to detect disturbances that may occur in the geomagnetic tail spontaneously, at some intermediate scales, in a wide range between the global scale of the entire magnetosphere and the characteristic ion scales in the plasma, namely the ion Larmor radius and the ion inertial length. In addition, the size of the grid cell, large as compared to those scales, – it is only twice smaller than the ion inertial length, – in principle excludes the possibility of detecting those small-scale current structures, which were discussed above. We point out again that those structures, if they arise in the course of evolution, play a key role in energy transformations in the system as a whole.

We believe that important and physically clear results can be obtained already in the one-dimensional simulation of the system containing the CS. First, an elongated, nearly one-dimensional CS is an appropriate model of the initial state of the system prior to the commencement of its rapid evolution. Second, as it was first pointed out by Petschek already in 1964, just restructuring ongoing at once on the long, almost one-dimensional segment of the CS (and not only in the vicinity of the emerging neutral line) is able to provide a high rate of transformation of electromagnetic energy stored in the system involving the CS, into energy of plasma flows, i.e. a high rate of magnetic reconnection. Third, we have a situation like in cosmic gas dynamics, where the thickness scale of shock fronts is much smaller than all the other scales, and this allows to regard them as one-dimensional discontinuities, specifying, however, the major constraints on the properties of large-scale flows that result from the conservation laws. In our case, the thin CS can be treated as one-dimensional structures, similarly determining the large-scale dynamics. Fourth, the computational resources available to us have allowed to carry out simulation quickly enough, but at the required high spatial and temporal resolution (at least a dozen cells on the ion inertial length), only in one-dimensional geometry.

We believe, and we have confirmed this by numerical simulations, that the decisive role in this one-dimensional dynamics belongs to the features of precisely the small-scale nonlinear kinetic structures, with thickness on the order of the ion gyro-radius, spontaneously occurring during the evolution of the system, as embedded in a thicker plasma layer. This complex dynamic picture appears in place of the original CS, that has formed during the previous slow, quasi-static stage of evolution.

In the numerical model, like in theoretical MHD models, the development of the restructuring process on the later stages is independent of the initial disturbance, the disturbance intensity gets spontaneously established at some finite level. Like in MHD, a finite amplitude disturbance appears having the form of a rarefaction wave that propagates in the background plasma from the original CS with the Alfvén speed. The rarefaction wave is followed with a slow disturbance which has a complex structure determined by kinetic effects. In this picture, usually one of two different characteristic forms of disturbance can be identified, corresponding to two types of nonlinear kinetic small-scale structures: the slow shock or the induced kinetic anisotropic CS. Simulation reveals that the longitudinal equilibrium being violated, without an external trigger, two fundamentally different evolution paths are possible, with emergence of one of these two structures, – for different values of the system parameters.

### 3.3. Nonlinear time-dependent problem: kinetic numerical model

For the numerical analysis, a simple one-dimensional (uniform along the $x$ and $y$ axes) model is used, its detailed description is presented in (Domrin and Kropotkin 2003). In the initial state, the system consists of a hot plasma, forming a CS of the Harris layer type (Harris 1962), and the surrounding homogeneous background plasma whose temperature is 10 times lower. The density of the background plasma density is equal to that of the hot one at $z = 0$. The dynamics of the ion (proton) plasma component is governed by the kinetic equation, whose solution is sought for by the particle-in-cell method. Electrons form a massless cold background.

It is proposed that the magnetic field remains flat during the entire simulation, $B_y = 0$, so that the self-consistent electric field $E_y$ and the tangential to the CS magnetic field component $B_t$ are determined by the only component of the vector potential $A_y = A(z,t)$. Fulfillment of the conditions being necessary for that, will be discussed later in Sec. 3.3.4. The fields $E_y$ and $B_t$ change along with the electric current density, which, in turn, is calculated based on the time-dependent ion distribution function. A non-zero magnetic field component $B_n$ normal to the CS is given at the initial moment and does not change in time. The boundary conditions at the origin $z = 0$ correspond to the central plane of the CS symmetrical over $z$ coordinate. Another boundary of the simulation box, $z = Z_0$, is located so that the size of the box far exceeds the thickness of the original CS, i.e. the initial plasma and magnetic field become almost uniform at that position. In our calculations, we took it equal to six characteristic scales of the initial CS. The electric field intensity $E_y$ at that boundary is calculated as the sum of the fields of two fast magnetosonic waves in a homogeneous plasma: one is propagating from outside towards the CS and is specified at that boundary, while the other is running in the opposite direction. Bearing in mind the relationship between the field $E_y$ and the disturbance $\delta B_x$ in each of these FMS waves, the boundary condition can be written in the form:

$$\frac{1}{c}\frac{\partial A(Z_0,t)}{\partial t} = 2E_i(Z_0,t) - \frac{4\pi V_A}{c^2}\int_0^{Z_0}\left[j(\varsigma,t) - j_0(\varsigma)\right]d\varsigma,$$

where $E_i(Z_0,t)$ and $A(Z_0,t)$ are the electric field of the incident wave and the vector potential at the boundary of simulation box at time $t$, respectively. Notations $j(z,t)$ and $j_0(z)$ refer to the current density at the point $z$ at time $t$ and at the initial moment, respectively, and the integration is performed over the entire simulation domain. Ion fluxes crossing the boundary from the outside and producing the time-varying contribution to the ion distribution function in the simulation domain, are calculated based on the motion of particles having the background concentration which occurs in crossed fields: the background magnetic field and the calculated electric field $E_y(Z_0,t)$.

Thus at the initial time there is an uncompensated magnetic tension in the system and its configuration is out of balance. Just from such non-equilibrium configuration, the phase of interest in the evolution of the system begins. The phase appears to be relatively slow compared to the duration of its start: as the field normal component $B_n$ is assumed small, the magnetic tension force is small and causes only a slow motion. Therefore one can speak of relaxation of non-equilibrium configuration which is given, "prepared" at the initial time, see Subsec. 3.1.2. In (Domrin and Kropotkin 2007a,b) the evolution of that non-equilibrium state is considered, both in the absence of external disturbance and in its presence. Here we restrict ourselves to the former case: the electric field $E_y$ and the convection velocity $v_z$ are equal to zero at $z = Z_0, t = 0$; the electric field of the magnetosonic wave impinging from the outside, remains equal to zero at all subsequent times.

### 3.3.1. The evolution involving formation of a pair of shocks

Since the evolution occurs under the action of magnetic tension proportional to the magnitude of the normal component $B_n$, the simulation was carried out for various regimes, which differ from each other by different values of the field component ratio, $B_n / B_0$, which varied within wide limits, $0,32 \geq B_n / B_0 \geq 0,04$. Here $B_0$ is the value of the magnetic field tangential component in the initial Harris layer away from the CS. In the simulation the dimensionless time $\Omega_0 t$ was used where $\Omega_0 = eB_0 / mc$ is the ion gyro frequency in the magnetic field $B_0$.

In Fig. 3.4 the evolution of the electric field $E_y(z)$ is shown, over the time interval $0,1 < t < 100$ for the ratio $B_n / B_0 = 0,2$. Here and below, the coordinate $z$ along the abscissa axis is in units of the Larmor radius $\rho_0 = v_T^{(h)} / \Omega_0$, where $v_T^{(h)}$ is the thermal velocity of hot ions. In the vicinity of the central plane a wave is generated, which propagates towards the boundary $z = Z_0$. Calculations show that its propagation speed is in the range of 0.9 to $1,1 V_A$, $V_A = B_0 / \sqrt{4\pi m N_0}$ is the Alfvén speed in the magnetic field $B_0$ in the background plasma of density $N_0$. Thus, the velocity of the wave propagation and its polarization as well, indicate that it is a fast magnetosonic (FMS) wave. It is seen that this result of kinetic simulation is consistent with the concept of the MHD theory of the plane discontinuity decay.

Dependence $E_y(z)$ in later times $t = 176, 208, 240, 272$, as well as the profiles of density $N(z)$, of the magnetic field tangential component $B_t(z)$, and of the components of the mean mass velocity, $V_x(z)$ and $V_z(z)$, are shown in Fig. 3.5 and 3.6. For higher resolution, only the half of the simulation domain closest to the central plane is shown in these figures. The plots show formation of the magnetic field jump, which is naturally interpreted as the front of a slow switch-off shock. Over time, the front moves with a velocity that lies in the range from $0,04\,V_A$ to $0,06\,V_A$. The velocity of the front motion according to MHD theory is similar but slightly higher, around $0,08\,V_A$. A better consistence with this value demonstrates the speed of the front, at which (as it should be on the switch-off shock) the plasma convective motion predominantly in the $z$ direction changes for the motion predominantly in the $x$ direction. The speed of its propagation is close to $0,07\,V_A$. In MHD this front must coincide with the shock front, as defined by the magnetic field jump. The difference of velocities obtained as a result of simulation, from those predicted by the MHD theory, is possibly due to the strong inhomogeneity of plasma in the zone of shock formation. The relation $\mathbf{E} + [\mathbf{vB}]/c = 0$ is approximately satisfied on both sides of the front. Thus, at the later stage of evolution, at times, far exceeding the ion gyro period, an analog of the time-dependent MHD disturbance of the Petschek solution type (Petschek 1964, Pudovkin and Semenov 1985) was obtained in our numerical model. And the strong decrease of $B_t$ behind the shock front can be interpreted as the effect of "dipolarization" during substorm activation.

Calculations at other, but not too small values of $B_n/B_0$ show that always in the vicinity of the CS central plane, at first a fast magnetosonic wave is generated which propagates outwards. The wave amplitude is nearly proportional to the value of the magnetic field normal component. Later, near the central plane at a distance of $z/\rho_0 \sim 6$, nearly at the same time (at $t \sim 60$) jumps of $B_t$ begin to form. First, a slow steepening of the profiles occurs on a scale of about three Larmor radii. At $t \sim 120$ the jumps start to evolve (at large $B_n/B_0$ evolution begins a little earlier). The evolution manifests itself in the slope steepening, along with their departing from the central plane. Profiles of the magnetic field tangential component $B_t(z)$ at the time $t \sim 208$, at $B_n / B_0 = 0,32; 0,24; 0,20; 0,16$ are shown in Fig. 3.7. Each of the jumps moves ahead by a few Larmor radii, and then decays. Production and decay of the jump is repeated several times. As shown above for $B_n / B_0 = 0,2$, in general terms the process follows the MHD pattern, but the very appearance of jumps on the Larmor radius (or ion inertial length) scale, as well as some finer features of the evolution should be regarded as purely kinetic effects.

Away from the central plane, where the slow disturbance has not reached yet, a gradual decrease of $B_t$ takes place. For larger $B_n$, the $B_t$ reduction is more significant. This behavior also has an analogue in the MHD: it corresponds to the fast self-similar rarefaction wave propagating ahead of the slow shock.

In accordance with the magnetic field variation, the induction electric field appears first on the central plane and then spreads with the Alfvén wave velocity as a FMS wave. In the region behind the front of this disturbance, first an increase of $E_y$ occurs, while the rate of growth is nearly proportional to $B_n$, and then the growth gradually ceases. That $E_y$ stabilization starts at the center of the layer, at $t \sim 120$, and at $t \sim 400$ it occupies already the entire simulation domain. Again there is an agreement with the MHD pattern, where the smooth $E_y$ growth occurs as far as the observation point is within the zone of self-similar rarefaction wave. After passage of its trailing edge, the electric field and the corresponding convection velocity in front of the shock remain constant and nearly proportional to $B_n$.

The plasma density disturbance is also roughly in line with the MHD model: the density increases in the slowly expanding area; this increase has the character of the density jump of the slow switch-off shock. However, there is a very approximate match, which is explained on the one hand, by the presence of hot plasma nonuniform over $z$, in our kinetic model, and on the other – by especially strong fluctuations of namely calculated density, in that part of the simulation domain.

*3.3.2. The evolution with formation of forced kinetic current sheet (FKCS)*

At small values of the $B_n / B_0$ ratio the initial stage – formation at the layer central plane of a fast disturbance escaping with the Alfvén speed, remains the same as described in Sec. 3.3.1, but the magnitude of the disturbance is smaller.

Regardless of the $B_n/B_0$ value, the duration of that stage is about $t \sim 100$. However, if $B_n/B_0$ is less than the critical value, which in our particular model is approximately equal to 0.14, the subsequent evolution follows a quite different path.

Spatial profiles of the magnetic field tangential component $B_t(z)$, of the electric field $E_y(z)$, plasma density $N(z)$, and average velocities $V_x(z)$ and $V_z(z)$ for $B_n/B_0 = 0,12$ at some stages of evolution are shown in Figs. 3.8 and 3.9. Instead of forming a slow shock at $t \geq 70$ at the central area, formation of a new, extremely thin CS takes place, with thickness of about an ion gyro radius. After this, the structure becomes nearly stationary, it exists during the whole subsequent time of simulation. It is the presently well known FORCED KINETIC CURRENT SHEET (FKCS) (Kropotkin and Domrin 1995, 1996, Kropotkin et al 1997), in which momentum balance along the CS is provided by the ion anisotropy.

The steady-state momentum balance requires a quite certain finite value of the electric field $E_y = E_0$, given by the condition (3.12)

$$E_0 + V_{A1} B_n / c = 0.$$

A front of anisotropy propagates from the central plane; at the same time it is the front of the plasma doubled density. Its speed is equal to $V_{A1} B_n / B_0$, i.e. it is the Alfvén velocity calculated with the magnetic field normal component $B_n$.

Calculations show that the steady-state electric field is somewhat less than the value determined by Eq. (3.12), $E_y \simeq 0,7 E_0$. The field with that intensity determines the plasma convection speed ahead of the anisotropy front, equal according to the calculation $V_z \simeq 0,072 V_{A1}$ (as opposed to $V_z \simeq 0,12 V_{A1}$ according to the theory), and its velocity along the field lines behind the front is $V_x \simeq 0,7 V_{A1}$ ($V_x = V_{A1}$ according to the theory). The ratio of these velocities, equal to $B_n/B_0 = 0,12$ according to the theory, turns out to be $V_z/V_x \simeq 0,10$ in the calculation. The anisotropy front, where the plasma density increases from $N = N_0$ to $N = 2,3 N_0$ (according to theory it should be doubled), is moving with velocity $0.08 V_{A1}$ (as opposed to $0.12 V_{A1}$ according to the theory), and its width is about 20 Larmor radii $\rho_0$.

Differences from the theoretical model here also may be due to the fact that the disturbance does not propagate in a uniform background plasma, but in the presence of a layer of hot inhomogeneous plasma.

As can be seen from Fig. 3.10, calculations for other values of the magnetic field normal component, smaller than the critical value ($B_n / B_0 = 0.14$, $0.12$, $0.08$, and $0.04$), show that in these cases also the formation of FKCS begins at the same time and proceeds at the same rate (taking the time interval $70 \leq t \leq 500$ from the start of simulation). The current density increases in FKCS approximately at the same rate at all $B_n / B_0$ values. However, the total current concentrated in FKCS, at smaller $B_n / B_0$ increases slowlier. Away from the FKCS, the magnetic field tangential component decreases with time and reaches a steady-state level. The difference between the original value of $B_t$ and the steady-state one becomes smaller with decreasing $B_n$. These properties correspond to the rarefaction wave propagating from the CS as predicted by the MHD theory (Domrin and Kropotkin 2004a, 2007b). The electric fields, being established as a result of this evolution, satisfy Eq. (3.12), regardless of the $B_n / B_0$ value.

We emphasize once again that the FKCS formation and all the related dynamics take place against the background of remaining, but slowly evolving current structure corresponding to the layer of hot plasma, which remains trapped in the vicinity of the central plane, like in the initial Harris type layer, i.e. spontaneous formation occurs of a small-scale structure embedded in a thicker current sheet.

Ion distribution functions have been obtained. They have generally complicated anisotropy. Examples of distribution functions are demonstrated in Fig. 3.11. The red bar denotes the value of the hot ions thermal velocity (it is equal to the Alfvén velocity outside the plasma sheet, which follows from the transverse momentum balance), the green one – that of the cold ions. The distributions show different anisotropy features: fast bulk flows (plot d), double-humped distributions of counter-streaming flows (plot c), patchy distribution formed by ions on their Speiser orbits near the central plane (plot b). Note that the latter has very close resemblance with the plots obtained by Chen and Palmadesso (1986). This is quite natural since the Poincaré maps presented there for ions arriving in the (prescribed, not-self-consistent) CS, oscillating about the central plane and then escaping, should involve that same patchy structure of the system phase space which is also seen in the ion distribution functions obtained by us.

Today a number of experimental results have been obtained which are well consistent with these simulation results, and thus provide experimental confirmation of the FKCS theory. As an example, some of them are shown in Fig. 3.12 and Fig. 3.13.

In the limiting case (initially equilibrium!), when the normal component is equal to zero, the evolution of the system is absent as it should be: the system remains in its initial state during the whole simulation time.

### *3.3.3. Physical causes and specific features of different modes of CS evolution*

In the system considered in Sec. 3.3.1 and Sec. 3.3.2, at the initial time the macroscopic ion flux is given by the current density in the Harris type layer and is $y$-directed. Therefore, an unbalanced force of magnetic tension at the initial moment is $x$-directed and produces acceleration along the $x$-axis only. Consequently, at $t = 0$ we have $E_y = \partial B_x / \partial t = \partial j / \partial t = \partial V_y / \partial t = 0$. Relaxation of the non-equilibrium state gives rise to the $E_y$ field growth from its initial zero value.

In the presence of a uniform cold plasma background, the $E_y$ disturbance propagates on both sides of the CS in the form of FMS waves. This causes the convective "raking" of the cold plasma to the central region: beyond the CS the $y$-momentum balance leads to the condition $E_y + V_z B_0 / c = 0$. But on the central plane the $z$-directed motion stops, i.e. the mean velocity normal component is always equal to zero there, $V_z = 0$. The simulation has shown that the zone of deceleration may have different structure depending on the parameters of the problem. In that zone, thin kinetic quasi-stationary structures arise, with thickness of the order of the ion gyro radius, which, depending on the $B_n$ value, can be either slow shocks, or forced kinetic CS.

Particle distribution functions in these structures and in their neighborhood have a complex anisotropy. At the shock front, along with particles that are already in non-adiabatic orbits and are accelerated by the electric field in the current direction, there are those ones that drift across the CS adiabatically. Shocks can evolve further and "overturn", which leads to multi-flow motion of the plasma particles. Conceivably in the numerical simulation this is manifested in breakdown of the thin CS and formation of trains of nonlinear slow waves followed by re-steepening and formation of thin fronts.

Which of the structures, the collisionless shock or the FKCS, arises and when this occurs, depends on the competition between two processes. On the one hand, the electric field in the CS leads to acceleration of ions on Speiser orbits and formation, outside the CS, of backward flow of fast ions escaping from the CS along the field lines, which effectively slows down the average motion towards the CS. This is the tendency to formation of FKCS. It manifests itself at small values of the drift velocity $V_z$. At large $V_z$ the "raking" of plasma is faster, leading to its accumulation and increased pressure in the central region. This causes "repelling" of the magnetic field from the central plane, i.e. an effective increase in the CS cross-scale and hence, reducing the role of ions occupying the Speiser orbits. This is the tendency to formation of a slow shock.

Our simulation shows that the result of this competition is really critically dependent on the $B_n / B_0$ ratio. When it is less than some critical value, the FKCS arises, but if it exceeds this value, then switch-off slow shocks appear, at which the magnetic field abruptly decreases several times.

In the current structures considered, over a small scale, of the order of the ion gyro radius, the transformation occurs of electromagnetic energy into kinetic energy of ion fluxes. In general, it occurs in any nonlinear current structure in plasma, including the Alfvén, rotational MHD discontinuity. However, in the FKCS, as well as at the front of collisionless shock, in the course of transformation "free" energy is generated: strongly anisotropic ion distributions with interpenetrating fast ion flows appear. Further on, the "free" energy should dissipate during relaxation of the distribution function through interaction of particles with the waves of plasma turbulence generated by the unstable ion distribution itself. That is the way nonlinear dissipative structures arise in a collisionless plasma. And in this way, the magnetic field "annihilation" takes place, which is a necessary constituent of fast magnetic reconnection.

We emphasize that in a collisionless plasma with magnetic field, nonlinear dissipative structures are supported by two different processes, separated in time and space. First, transformation of electromagnetic energy into kinetic energy of particles and formation of unstable ion distributions takes place, and then – dissipation associated with damping of plasma waves excited by instability.

The initial energy conversion occurs in thin CS's, where the electromagnetic energy is converted into energy of ions flows accelerated by the electric field inside the CS (in the case of collisionless shock, there is a group of ions, which oscillate at the wave front and are accelerated in the electric field direction, see e.g. (Sagdeev 1964)).

As to the spatial scale of the relaxation process for unstable ion distributions, it depends on the processes of wave – particle interaction, actually operating outside the CS (for the shocks – behind the front). Those processes, on their nonlinear stage, are still not very well known, and this provides uncertainty with respect to the specified scale. In the particle-in-cell numerical simulation, that uncertainty is compounded by sampling effects and round-up errors, which lead to further uncontrolled numerical damping. It might seem that all this uncertainty fully hinders understanding of the basic features of nonlinear dissipative structures. However, this is generally not the case. It is known that in the case of (subcritical, laminar) collisionless shock, the major characteristics: entropy generation rate and the rate of energy transformation, – do not depend on the internal (oscillatory) shock-wave structure. The wave train behind the front can be longer or shorter depending on the properties of wave damping, but those major characteristics are determined only by the asymptotic values of the parameters of plasma and magnetic field on both sides of the front, away from it. Similarly, in the case of the FKCS, the rate of energy transformation depends only on the asymptotic characteristics. We see that for our purposes rather crude modeling is sufficient.

### *3.3.4. Discussion: Features of the system evolution in different formulations of the numerical simulation problem*

Our model is consistent with that approach to magnetospheric dynamics, which shifts the emphasis from the magnetic reconnection as a topological restructuring to the processes of local transformation of energy and its dissipation due to the emergence of extremely thin current sheets in the system. In a series of papers, which opens with an article (Birn et al 2001), various methods of numerical simulation have been applied to processes occurring in the vicinity of the magnetic field neutral line in the geomagnetic tail, in the absence of collisions. In the final paper of this series (Kuznetsova et al 2001, p.3809), following the earlier works (Galeev et al 1978, Biscamp et al 1995, Hesse et al, 1999), an important conclusion is made: the reconnection rate at the neutral line is itself determined by the large-scale (in our case – mesoscale) dynamics of the system, adapting to it. Namely the characteristics of that dynamics are clarified in our approach. The above rate is determined by the rate of energy transformation which, in turn, is given by the flux of electromagnetic energy towards the CS,

$$P = c \frac{EB_0}{4\pi} = \frac{B_0^2 B_n}{(4\pi)^{3/2} \sqrt{\rho_1}}.$$

As already mentioned, attempts of a more complete simulation of the "forced" reconnection process in the geomagnetic tail, not limited by the neighborhood of the neutral line, have been also made; a brief overview is contained in (Pritchett 2005). The latest in the series of these works involve three-dimensional simulation; the particle-in-cell method is applied to behavior of both components: the ions and the electrons. The electronic effects are carefully studied and considered to be of paramount importance. Those are electronic currents, electrostatic polarization, the ultra-thin electron sheet, electron acceleration to relativistic energies. However, such simulations are still unable to raise the ion/electron mass ratio above $m_i / m_e = 100$, i.e. this ratio almost one and a half orders of magnitude differs from the true one. Thus the role of electronic effects may be greatly exaggerated.

Another concern relates to the inclusion of "forcing" convection electric field. In recent models, just as is done here, it is turned on at the boundary of the simulation domain over $z$. But in contrast to our approach, a wave emanating from the simulation domain is not considered. And this is only true for a very short time, until the FMS wave from the central plane reaches the boundary of the simulation domain. In addition, the electric field is switched on right away as localized over $x$ But at location where the field is the maximum, there is a reverse of $B_z$, i.e. a neutral line is formed. Most research focuses on emerging magnetic reconnection. Unmagnetized ions and electrons, and therefore also the corresponding collisionless dissipation, appear. The paper (Pritchett 2005) specifies the complex picture of processes in the vicinity of the neutral line. Estimation of the reconnection rate is possible only at a certain initial phase, until the disturbance has not reached the boundary of the simulation domain. The above limitations of the approach, associated with formulation of boundary conditions, casts doubt on the validity of the results. Completely unexplored remains the process of magnetic field annihilation at large scale over $x$, in those parts of the CS remote from the neutral line. And that is, as we have emphasized, namely the process which provides fast energy transformation in the CS.

We have shown that the nature of this crucial process can be successfully figured out with a simple and physically more transparent one-dimensional kinetic model.

Solving one-dimensional non-stationary problem of the evolution of non-equilibrium CS in collisionless plasma with a nonzero field component $B_n$, is also the subject of a number of papers (both in the MHD approximation and in kinetics), for example, (Heyn et al 1988, Semenov et al 1992, Lin and Lee 1994, 1995, Hoshino 2005). Among them, a significant part relates to analysis of situations with a non-zero "guiding" field $B_y$, as well as of systems with asymmetric properties of plasma and field intensity on both sides of the CS. The results of corresponding numerical simulations are highly diverse.

We confine ourselves, however, with the symmetric case, $B_y = 0$. Symmetry is typical for the CS in the geomagnetic tail, and the condition $B_y = 0$ at small $B_n$ provides an essential role of non-adiabatic ion trajectories. In the kinetic consideration, the latter means that processes of ion acceleration can be efficient in the system, and consequently also the process of magnetic reconnection can – in the sense of efficient transformation of electromagnetic energy into energy of ion flows.

For the same symmetric case with $B_y = 0$ and for the slow switch-off shock as a basic element of the reconnection process, numerical simulation was carried out before (Omidi and Winske 1989, Winske and Omidi 1990, Vu et al 1992, Lin and Lee 1994, 1995, Hoshino 2005). In those studies effects were found that are absent in our results, and conversely, the most important effects found out in our studies – FKCS formation, ion anisotropy forms – were not recorded in those papers. This requires further analysis.

In solving one-dimensional non-stationary problem of the evolution of non-equilibrium CS with non-zero field component $B_n$, carried out without any additional restrictions, our description of the field with the only nonzero component of the vector potential $A_y(x,z,t)$ is generally speaking, insufficient. In fact, with such a description, for all $t$ the magnetic field $y$-component is absent, $B_y = \partial A_x / \partial z - \partial A_z / \partial x = 0$, and this is generally not the case.

Indeed, when the electric field is non-zero, $E_y \neq 0$, an electron drift appears,

$$\mathbf{v}_e = c\frac{[\mathbf{E}\mathbf{B}]}{B^2},$$

producing the corresponding electron contribution to the current density, with nonzero $x$- and $z$-components. The electron current, generally speaking, is not fully compensated by the ion currents over $x$ and $z$, since the ions, in contrast to electrons, can be magnetized not everywhere. Then the magnetic field $y$-component $B_y(z,t)$ is generated, although in the initial state this component has been zero everywhere. This is followed, in turn, by generation of $x$-component of the induction electric field, and thus an oscillatory elliptically polarized wave train arises. This effect is observed in numerical simulations in (Lin and Lee 1994,1995) and in the earlier works (Omidi and Winske 1989, Winske and Omidi 1990, Vu et al 1992).

However, the effect can be much suppressed and be ignored. For this to occur, the ions should have the same average velocity over $x$ and $z$ as the electrons. This is really possible under certain restrictions on the temporal and spatial scales of the $E_y$ field variation.

In the ion motion in the adiabaticity domain, its average velocity in the $xz$ plane is determined by the electric drift only, and coincides with the electron velocity so that the current density components over $x$ and $z$ are zero in that domain.

Note that in the two-dimensional geometry of the magnetic field, in the charged particle motion the generalized momentum $P_y = p_y + A_y/c$ is conserved.

We also first assume that $B_y = 0, E_y = 0$. Then the field lines lie in the planes $y = \text{const}$, being iso-contours of the vector potential $A_y$, and the projection of the particle trajectory on the $y = 0$ plane oscillates, located between the two field lines $A_\pm = c\left(P_y - p_{y\pm}\right)/e$. Here $p_{y\pm} = \pm\sqrt{2m\varepsilon}$, $\varepsilon$ is the particle kinetic energy. Thus the average velocity of motion across magnetic field line is the same for electrons and ions and is equal to zero.

Now let the magnetic field depend on time, and then there is a nonzero alternating induction electric field. Consider a layer with current having thickness on the order of the ion gyro radius $\rho_i$ in the field $B_0$; adiabaticity is violated in the layer. Assume that this layer and the associated magnetic field jump move slowly over $z$, with velocity $\sim V_{A0}\mathrm{O}(\varepsilon), \varepsilon = B_n / B_0$. This is the case with the switch-off shock front in the MHD model, e.g., (Semenov et al 1992). In the kinetic numerical model that can be verified *a posteriori*. Under this condition, the electric field is almost unchanged on the scale of the layer: the relative change of $E_y$ is of the order $\varepsilon$. Then up to accuracy $\mathrm{O}(\varepsilon)$, the electric field $E_y'$ is zero in the reference frame moving in $x$-direction with velocity $v = cE_y / B_n$.

However, in order to be able to neglect the variation of the field $E_y$ on the non-adiabatic part of the ion orbit, it is also necessary that the field has changed only slightly over the entire time interval when the ion is in that part. The duration of that interval is estimated as $\rho_i B_0 / v_{Ti} B_n = \Omega_{in}^{-1}$ – it is of the order of the ion gyro period in the field $B_n$.

It is seen that if the characteristic time scale $\tau$ of the field $E_y$ variation is large, $\Omega_{in}\tau \gg 1$, then there exists a reference frame where the average velocity over $x$ of both electrons and ions, with accuracy $\mathrm{O}(\varepsilon)$ and $\mathrm{O}(1/\Omega_{in}\tau)$, is equal to zero. It follows that in this layer the current density turns out to be almost zero.

The specified extended time scale may be estimated as follows. In the equilibrium steady state all the plasma moves in the $x$ direction at an average velocity equal to the Alfvén velocity calculated with the density of cold plasma $\rho_1$, i.e. $V_{A1}$. Based on the Euler equation, the time which is needed to accelerate plasma up to this value in the central part of the CS, is estimated as follows:

$$\rho_0 V_{A1} / \tau \sim \frac{B_0 B_n}{4\pi L_z},$$

where $L_z$ is the scale of the initial CS, so that

$$\tau \sim \frac{B_0}{B_n} \frac{V_{A1}}{V_{A0}} \frac{L_z}{V_{A0}} = \frac{B_0}{B_n} \frac{V_{A1}}{V_{A0}} \frac{L_z}{v_{Ti}},$$

where we have taken into account that for the equilibrium CS, $V_{A0} = v_{Ti}$. Note that $v_{Ti} = r_{Li} \Omega_{in} B_0 / B_n$. Then we obtain:

$$\Omega_{in} \tau \sim \frac{V_{A1}}{V_{A0}} \frac{L_z}{r_{Li}}.$$

As in the initial period the CS thickness exceeds the Larmor radius $r_{Li}$, then $\Omega_{in}\tau \gg 1$. This is true even for extremely thin ion CS, with $L_z / r_{Li} \sim 1$, if we consider the background plasma density low, so that $\rho_1 / \rho_0 \ll 1$ and respectively, $V_{A1} / V_{A0} \gg 1$.

It is seen that our approach is valid if the problem has two small parameters: $B_n / B_0 \ll 1$ and $\rho_1 / \rho_0 \ll 1$. But this is exactly the approximation adequate to the CS in the geomagnetic tail, see Sec. 3.1.2.

Like in the MHD model, in the fast magnetosonic disturbance running ahead of all the others, the electric field propagates over $z$ on a small time scale, setting in $x$-directed motion the plasma in the central part of the system, and causing, outside the original CS, the convective plasma raking towards the CS on its both sides. Such a raking results in forming of a high density layer which expands slowly, over a long time scale. The time scale appears to be extended, first because of the fact that acceleration of the hot plasma forming the CS initially, requires a long time if its density is high. Second, the raking of the background plasma results only in a slow rise of central density if the background plasma density is low. The compression wave is nonlinear, its front is steepening, and the shock forms. This disturbance still continues to spread slowly. Therefore at its front which is a thin current sheet formed by non-adiabatic ion motion, the electric field undergoes only a small change which may be neglected. As a result, the $x$-directed electric current is not formed, and no magnetic field component $B_y$ appears.

A similar argument is also suitable for the case when not the shock but the FKCS is formed.

Turning to the conditions of our numerical experiment, confirmation of the fact that the ions move in the $xz$ plane along with the electrons, and conditions for generation of the $B_y$ field do not appear, might be obtained *a posteriori* from the simulation results. It suffices to compare the evolution of the electric field, the magnetic field, and the average ion velocity components. We have carried out such a comparison. We have found out that the profiles of the field $E$ and of $\|[\mathbf{vB}]\|$ change relatively slowly, and in fact they are the same, up to the errors of the numerical experiment.

So it should be concluded that evolution of the CS can proceed over highly different paths, as it is really evidenced by different results of numerical simulations. In the papers (Lin and Lee 1994, 1995) and in our work they have been carried out on similar mathematical patterns, however, for markedly different initial conditions.

The initial conditions adopted by us, have serious physical justification. The current sheet is formed quasi-statically, as a layer of hot plasma separating two regions with oppositely directed magnetic fields in which there is only a background plasma with $\beta \ll 1$, and with temperature and particle densities much lower than in the CS. This is a "residual" population which either remains there or appears from an external source (the ionosphere in the case of the geomagnetic tail), after "raking" of plasma into the plasma sheet which occurs on the quasi-static phase, see Sec. 3.1.2.

In the conditions emerging in this way, $V_{A1}/V_{A0} \gg 1$, we see that evolution occurs without generation of the transient field $B_y$ and thus without a train of oscillatory electromagnetic disturbance behind the front, escaping from the central plane.

We also note that we have studied the configuration, in general, quite different from that which usually is treated as a solution to the Riemann problem (Heyn et al 1988, Semenov et al 1992, Lin and Lee 1994,1995). A layer of hot and dense plasma is not set initially there, but appears as a dynamic structure, behind the compression jump – the switch-off shock: it is formed from the cold background plasma crossing the wave front. Therefore, an appeal to the well-known in the MHD Riemann problem is only intended to roughly outline the possible course of evolution of the system under the action of uncompensated magnetic tension: the spread of a slow disturbance from the central plane, which maintains the conversion of magnetic energy into energy of the plasma flow. Numerical simulation shows, however, that even on a qualitative, descriptive level it may occur in a substantially different manner. That is, when the FKCS is formed, the conversion does not occur inside the propagating front, but right in the central region of the CS. Prior to that, there occurs a gradual, extended in time acceleration of all the plasma in the plasma sheet, rather than adding through the wave front of fresh plasma, at once with velocity equal to the Alfvén one.

### 3.3.5. The relationship of different types of disturbances in the plasma sheet. A scheme of the event sequence

In the recent years, in experimental data obtained from spacecraft in the geomagnetic tail, nearly at the same time non-linear time-dependent structures of two quite different types were discovered. On the one hand, very thin current sheets sporadically appear in the plasma sheet, embedded in a thicker plasma structure, e.g. (Runov et al 2006). On the other hand, in that same region, on intermediate scales of a few $R_E$, nearly chaotic strong variations of the magnetic field and associated plasma motions are observed, e.g. (Borovsky and Funsten 2003, Weygand et al 2005). They form some sort of turbulence. In relation to that turbulence, a number of unresolved questions exists. What does drive it? What is the energy source? In observations, bursty bulk flows (BBF), see e.g. (Sergeev, 2004) may be the proper candidate. But what is their origin?

From the above, it may be conceived how could these two types of disturbances be interconnected and form a unified pattern of the specific turbulence in the plasma sheet (Kropotkin and Domrin 2009a,b).

The magnetic field restructurings occurring in the geomagnetic tail at small time scales $T_1$, can cause violation of the longitudinal equilibrium in the plasma sheet in areas with size $L$ satisfying the condition (3.9). Because of the $B_n$ smallness, relaxation that follows should occur on a much larger time scale $T_2 \gg T_1$. Along with the embedded thin current structures, ion flows of two different types appear: first, the flow outgoing from the CS along the field lines with twice the Alfvén speed, and secondly, the flow of ions accompanied by electrons, remaining inside the CS and in its immediate vicinity and moving with the Alfvén speed Earthward (or anti-Earthward) – a fast plasma flow. Since these fast flows occur on a relatively large time scale $T_2$, themselves and their effects are mainly of *quasi-stationary* nature. In the observations they may appear, first, as quasi-steady longitudinal fast ion flows (beamlets) in the auroral magnetosphere, and second – as quasi-stationary convection jets, with the corresponding electric and magnetic fields. Since the disturbed zone of the CS where the relaxation takes place, is finite on its scale over the $y$-direction also, exactly jets should appear within the CS. They correspond to quickly shrinking magnetic flux tubes, "dragged" through the adjacent ones which are in nearly equilibrium, steady state, see Fig. 3.14.

However, the effects being considered have fast wave constituents as well. Fast MHD rarefaction waves escaping from the CS, while propagating on scales exceeding the disturbance wavelength, $\geq V_{A1}T_1 \gg L$ (see Eq. (3.5)), transfer the disturbance electric field $E$ to the vicinities of the initial relaxation zone. This is followed there by thinning of the CS which occurs in a *quasi-stationary way*: its rate at a given point grows relatively slowly, following the change of $E$ which occurs on the time scale $T_2$. Locally, this thinning can generate a new non-linear tearing instability and the related process of fast reconnection in a new limited area. Further developments should follow the scenario described above. The picture may be complemented and complicated by fast MHD disturbances coming as waves reflected from the ionosphere.

As we have shown, the formation of thin kinetic CS is accompanied by local generation of fast ion flows. The latter, apparently, can in their turn maintain a turbulent cascade of decreasing scale non-stationary two-dimensional vortex motions in the plasma sheet, and the associated magnetic field variations. These slow motions with the corresponding electric fields are no longer connected with wavelike MHD disturbances. Therefore, at this level the plasma vortex motions and thin CS appear to be "decoupled". Note however, that it is the size of the vortices that can determine the scale $L$ which characterizes the size of the regions of longitudinal equilibrium violation, of embedded kinetic CS appearance, and of fast plasma flows. In fact, the vortices and their associated magnetic variations should lead to inhomogeneities in the thickness of the plasma layer having dimensions in the CS plane of the order of a few $R_E$. And on the thinner areas with a lower $B_n$ value, violations of the longitudinal equilibrium are easier to occur.

Thus, we arrive at the following scheme of successive events at medium distances in the geomagnetic tail, where the dynamics are viable involving neutral lines formation, and magnetic reconnection.

• A local thinning of the TS takes place.

• Locally, the threshold of the linear tearing instability (calculated for a system *frozen* at each successive moment of slow, quasi-static evolution) is reached and then is exceeded. (Or combined instability, see Subsec. 2.4.4 and 2.4.5, comes into play.)

• The dynamic bifurcation occurs – a quick breakdown of local equilibrium with formation of a neutral line (NL).

• In the vicinity of the NL a pulse of electric field is formed.

• The pulse propagates over the tail lobes as a MHD *disturbance*. The reason is that the currents of the initial disturbance are concentrated in a finite region over $x$, and so, away from the CS, the disturbance is not evanescent, exponentially decreasing, – like for tearing disturbance in the one-dimensional system, but is propagating, escaping from the CS.

• The localized MHD disturbance involves a pair of field-aligned currents – away from the CS it is the Alfvén wave. The reason for this is that the initial disturbance currents are concentrated in a finite region over $y$ as well.

• Near the CS this disturbance raking up the plasma to the central plane, has the character of a fast magnetosonic wave.

• In a *short Alfvén time* $T_1$, away from the NL in the vicinity of the CS, an increase of the tangential field is formed (and an additional thinning of the CS occurs).

• The CS becomes slightly out-of-balance there.

• On significantly longer times $T_2 \gg T_1$ the plasma flow is generated along the CS, a new, relatively slowly growing electric field appears, the evolution leads to formation of the FKCS or of the shock pair.

• Inside the CS, the arising plasma flow, reaching gradually the Alfvén velocity, results in the vicinity of NL to the relatively slow, on time scale $\sim T_2$, rise of the reconnection rate that occurs in agreement with the simulation by Kuznetsova and Hesse. The limiting achievable reconnection rate corresponds to spreading of the plasma along the CS with the Alfvén velocity – in line with the establishment of the FKCS regime or that of the shock pair.

- Outside the CS, the disturbance with a relatively slow increase of electric field, propagates to both sides of the CS as an MHD rarefaction wave.
- Relatively slow, on time scale $\sim T_2$, motions outside the CS directed towards it, occurring as a result of local, mesoscale restructuring, cover also some areas away from the area of the FKCS or the shock pair formation. The corresponding disturbances are transported there by the MHD signal from the latter area, and occur as quasi-stationary there, since the Alfvén time is always small, $T_1 \ll T_2$.
- On some *new* sites local CS thinning takes place.

Then the sequence of events is repeated.

The locally emerging picture is *quasi-one-dimensional*. As a whole, the evolution of flows and electric fields is determined by large-scale *three-dimensional* dynamics of the configuration. Therefore, in the results of *global modeling* (Kuznetsova et al 2007) there are parts of the plasma sheet moving with a variety of speeds and therefore with different fields $\mathbf{E}$. However, every such area should be a *reconnection layer*: at every point on the $(x, y)$ plane there is a sink (or source) of electromagnetic energy; energy is exchanged between the plasma flow and the magnetic field. Since in this case $E = -VB_n/c$, i.e. $B_n \neq 0$, it can be said in other words that at every such point the magnetic tension $B_t B_n / 4\pi$ does work. The process of the energy transformation is stretched on a large surface of the reconnection layer, which ensures its high speed.

To accommodate these circumstances, in the global simulation (Kuznetsova et al 2007) the MHD code is supplemented with electric fields accounted for, which are estimated using some artificial scheme in the vicinity of the neutral lines arising. This results in very significant and impressive effects. The global quasi-periodicity appears in the system dynamics: periods of accumulation of magnetic energy in the geomagnetic tail alternate with periods of discharge of that energy in the processes of magnetic reconnection. Thus the model becomes able to reproduce the substorm cycle.

# 4. Concluding remarks. Prospects for further research

For a given equilibrium in the magnetoplasma system involving a current sheet (the field $E = 0$), its tearing instability means that an arbitrarily small perturbation of the magnetic field grows exponentially. Hence a nonzero electric field appears, $E_y \sim e^{\gamma t}$. Moreover it appears everywhere, including the emerging null lines.

When considering non-linear tearing instability – the dynamic bifurcation, it is understood that the system has previously experienced quasi-static evolution, i.e. it passes through a continuous series of quasi-equilibriums, so that it involves a nonzero but small electric field $E_y \neq 0$. Then there occurs, in the qualitative sense, the same as in the linear tearing instability. Only the fast growth of disturbance no longer follows the exponential law, there is a time interval where the process is described as the "explosion" instability.

But in addition, at the end of the nonlinear dynamical process of bifurcation, there is the passage of the system to a new quasi-equilibrium, so that the field $E_y$ is again small and it is again determined by the rate of quasi-static evolution, like before the equilibrium breakdown – the dynamic bifurcation.

In the generic situation, in the model of dynamic bifurcation, the rate of quasi-static variation of the (mesoscale) magnetic configuration is given by the parameter $\Gamma$. The value of this parameter which has existed before the dynamic bifurcation, before the neutral line appearance, remains after the dynamic bifurcation as well, when the neutral line already exists. As follows from the theory, this parameter, along with the coefficient $\kappa$, as well as the value of $\Theta$, determined by the value of a small initial disturbance, sets the characteristic duration $\Delta t$ of the equilibrium breakdown, its delay $T$, as well as the new value of the field $E_y$. Just that new field $E_y$ value determines the magnetic reconnection rate at the newly emerged neutral line. Now this field is nonzero on the CS central plane also, the neutral line included, in contrast to the evolution prior to the equilibrium breakdown, when that field being given by the same parameter Γ value, was nonzero outside the CS central plane only, and resulted in thinning of the CS.

*After* the dynamic bifurcation that generates topological jump of the system to a new equilibrium involving a neutral line, the following features may be pointed out: (a) in the vicinity of the emerged neutral line, there is a quasi-stationary configuration of the type that appears in the numerical simulation (Kuznetsova et al 2007); (b) away from the neutral line, there is a reconnection CS (reconnection layer) of one of two possible types: either the pair of switch-off shocks or the FKCS. It is at this thin CS that the energy transformation basically occurs.

What is the global pattern of activity in the system after the occurrence of a slow CS thinning accompanied by accumulation of magnetic flux in the system? That pattern should be a set of reconfigurations spaced apart in time and space – bursts of magnetic reconnection at the emerging neutral lines, combined with formation of medium-sized (in the CS plane) reconnection layers. Plasma flows accelerating on those reconnection layers (under action of the new $E_y$ fields that arise after the dynamic bifurcations) serve as those agents which carry the effect of restructuring on the surrounding medium-sized sites and breeds there new fast balance breakdowns.

Such an intermittency pattern is reminiscent of, for example, the boiling process in liquid. The effect of the total heating of the liquid, with an increase of its temperature up to the boiling point, changes for a process of vaporization, when the input heat is passed not to liquid but to vapor. Vaporization occurs through formation of individual bubbles of vapor. Their buoyancy generates additional convective fluid motion; that is accompanied by the emergence of new local sites that are preferred for the formation of bubbles.

In this paper, the complex dynamic processes resulting in fast conversion of energy in the magnetospheric collisionless plasma which include cycles of accumulation and discharge of electromagnetic energy, fast processes of equilibrium loss, embedded thin CS, mesoscale turbulent plasma motions, and strong variations of the magnetic field, are considered from a unified point of view. It is argued that the central role belongs to nonlinear dynamic processes of thin kinetic CS formation. Those structures are *quasi-stationary*, and they facilitate the fast conversion of electromagnetic energy into energy of plasma flows. The latter, in turn, can be a source of energy for the two-dimensional hydromagnetic turbulence in the plasma sheet. The resulting intermittency, apparently, is a characteristic feature of turbulent disturbances, experimentally observed in the plasma sheet (see, e.g., (Klimas et al 2004) and references therein).

The global pattern of activity in the system outlined above, although based on a set of separate partial results presented in this work, is still only a scheme to be thoroughly verified. The verification appears to be possible to implement in the future in new improvements of the global MHD model, as a further development of those improvements which have already been done in the paper (Kuznetsova et al 2007) and its sequels.

In general, among the important results obtained in the last decade, we would like to highlight a series of numerical studies on the GEM Challenge program where the basic features of magnetic reconnection occurring in the vicinity of the neutral line have been identified. At the same time those studies have strongly supported the idea that the reconnection rate is determined by the conditions that exist outside the vicinity of the neutral line. And they were the basis for global modeling of 2007, which, as has been said, demonstrates the quasi-periodic intermittency.

What new additional research could be undertaken in the development of that modeling of 2007?

First, the problem remains unclear in the 2007 paper as how do those neutral lines emerge, the finding of which lies at the beginning of the simulation algorithm. In the MHD simulation being used, the emergence of the field $B_n$ reversal to produce such a domain with the neutral line, is impossible. It is not possible unless there is *tearing instability*. Does the instability really arise in the model – because of numerical dissipation? Does a fast nonlinear breakdown of equilibrium occur, with formation of finite amplitude tearing disturbances? Isn't it possible to examine "under a microscope" what is going on?

Second, whether or not it is possible in this modeling to identify those areas of the CS and the (relatively short) periods of time where and when the bursts of electric field and the emergence of a fast plasma flow occur – after a burst of reconnection (non-linear tearing "explosion"?) at the neutral line of the magnetic field? Is that generation of the reconnection layer essentially close to the one-dimensional models presented in this paper?

In general it is of fundamental importance here that the dynamics of the geomagnetic tail should involve *sporadic* action of a certain fast mechanism which results in dissipation of magnetic field energy, converting it into energy of plasma flows and in heat. This occurs in those regions where $(\mathbf{jE}) > 0$ – in the regions of *reconnection layers*, i.e. in those regions that are (a) occupied by a current sheet (CS), and in which (b) the electric field has penetrated into the CS. For this reason, the central role should be assigned to non-linear dynamic processes of *formation* of thin kinetic CS associated with the *appearance* of the induced electric field. The result is the embedded nonlinear current structures of the *reconnection layer*. The attempt looks promising to test these theoretical ideas on the numerical model, in which now *really exist both thin (almost one-dimensional) current sheets and elements of fast sporadic dynamics*.

*Acknowledgements.* This work was supported by the Russian Foundation for Basic Research (project no. 09-05-00410).

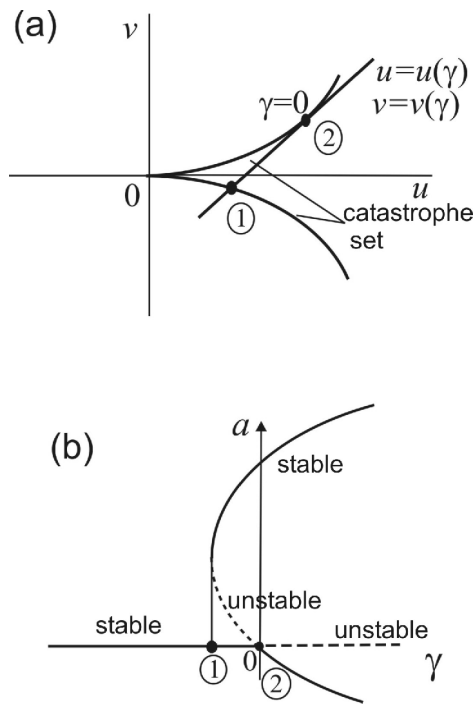

Fig. 2.1. The equilibrium state bifurcation on the evolution path (a) in coordinates $(u,v)$ and (b) in coordinates $(a,\gamma)$.

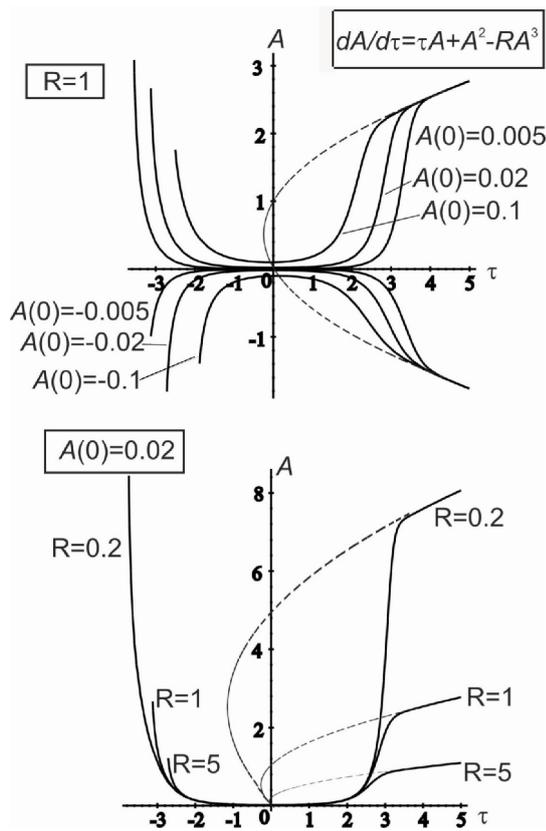

Fig. 2.2 Integral curves.

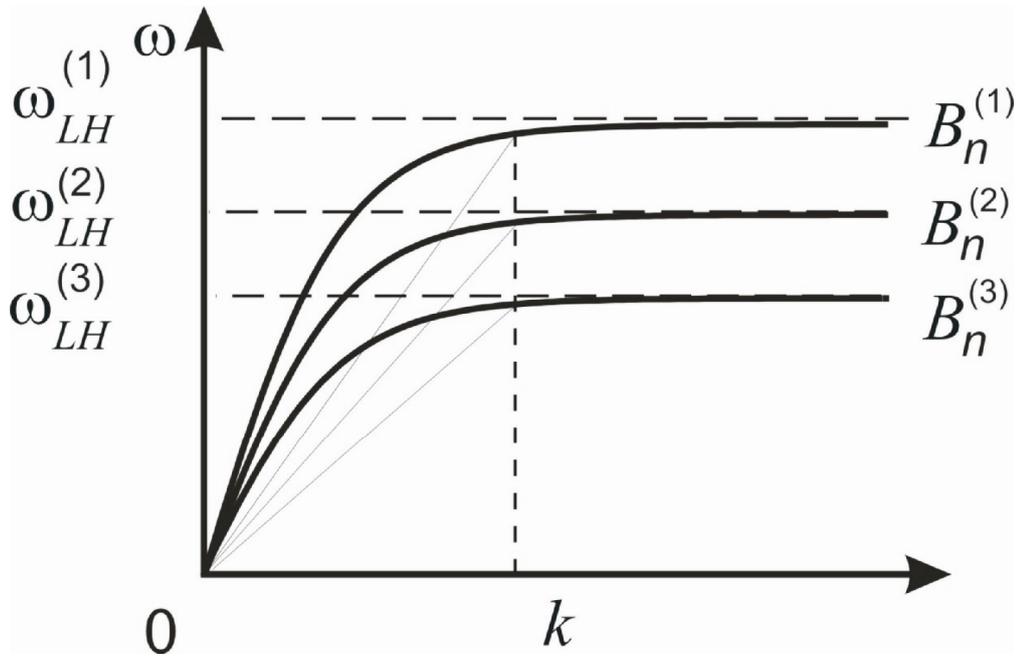

Fig. 2.3. Dispersion curves for different $B_n$, $\omega_b^{(1)} < \omega_b^{(2)} < \omega_b^{(3)}$.

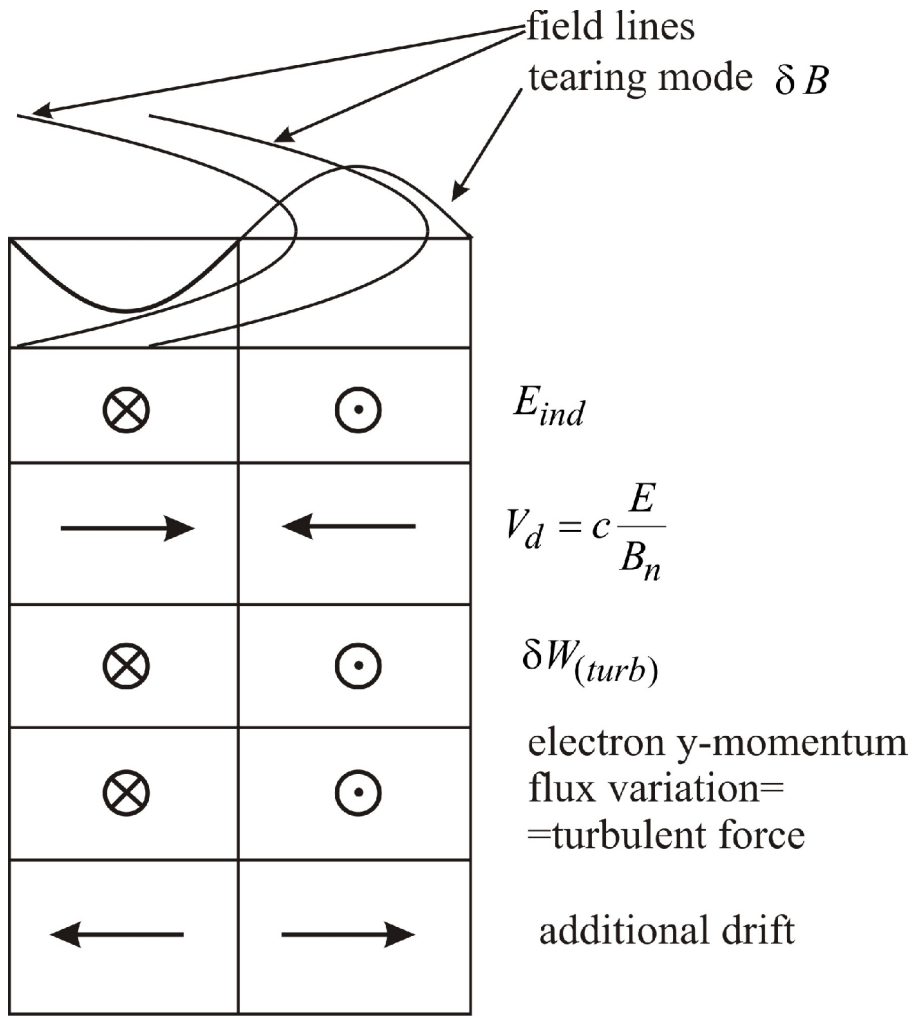

Fig. 2.4. Disturbance characteristics involved in the feedback loop.

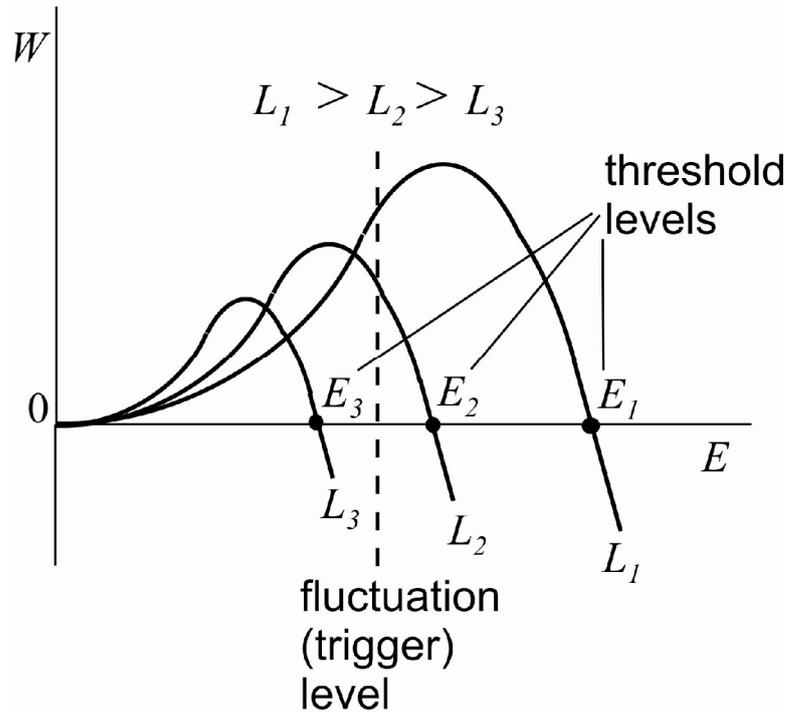

Fig. 2.5 Equilibrium loss depending on CS thinning.

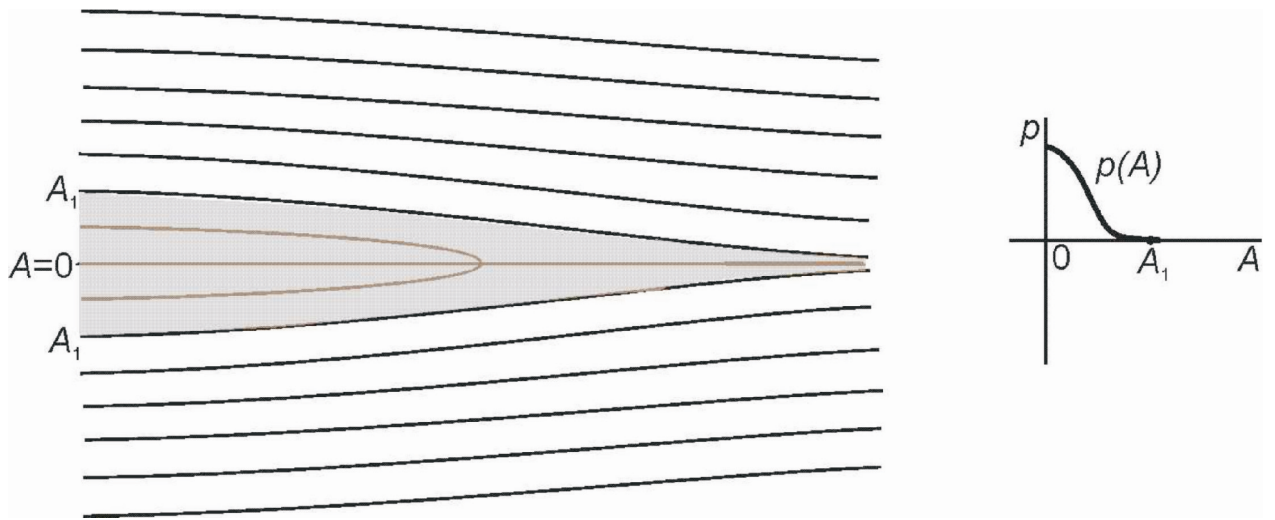

Fig. 3.1. Stretched configuration with a plasma sheet. Pressure dependence on the vector potential $A$ is shown. $A = \text{const}$ on the field lines.

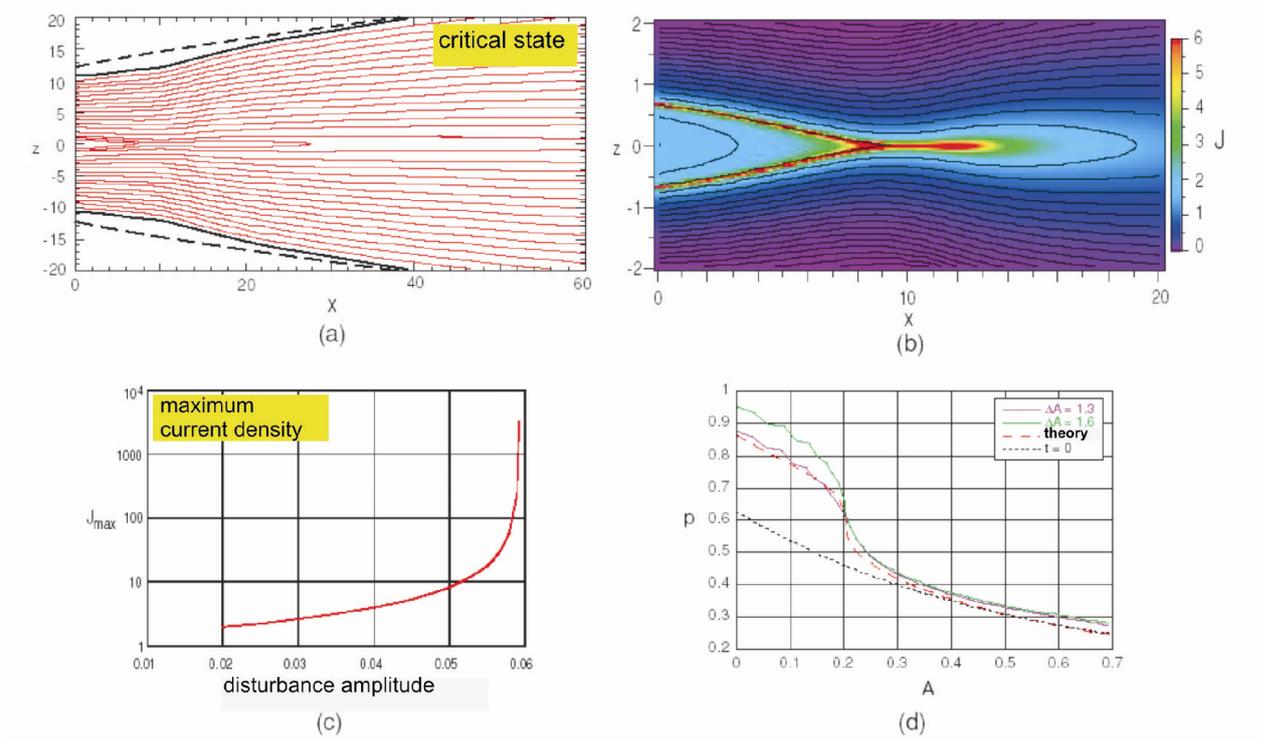

Fig. 3.2. (a) Critical state arising when the boundary is adiabatically deformed (solid black lines). The dashed lines show the initial state. (b) The inner region zoomed in; the current density is color-coded. (c) Maximum current density as a function of the boundary deformation magnitude. (d) Pressure profile $p(A)$ for the magnetospheric configurations corresponding to boundary deformations shown in figures (a) – (c). The dotted black line depicts the initial dependence, and the red dashed one – that dependence in the critical state calculated in the adiabatic theory. Two solid lines show the final results of time-dependent MHD simulation for two disturbance magnitudes close to the theoretical limit. (Following the results of (Birn and Schindler 2002) and (Birn et al 2003)).

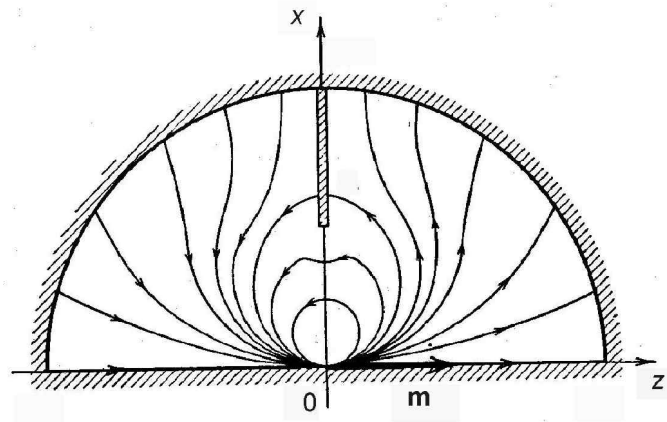

Fig. 3.3. Two-dimensional model of the geomagnetic tail (Somov and Syrovatskii 1974). The field is formed by the two-dimensional dipole $\mathbf{m}$ and the currents on the boundaries including (a) a segment of the $x = 0$ axis, (b) half-circle with radius $R \to \infty$, and (c) a segment of straight line on the $z = 0$ axis corresponding to the infinitely thin CS. The corresponding boundary conditions have the form: $A_{(a)} = 0$, $A_{(b)} = C_1 \sin \varphi$, $A_{(c)} = C_1$. The angle $\varphi$ is counted from the dipole $\mathbf{m}$ direction. The dipole-type singularity at the coordinate origin $r = 0$ has the form $A(r,\varphi) \to m \sin \varphi / r$.

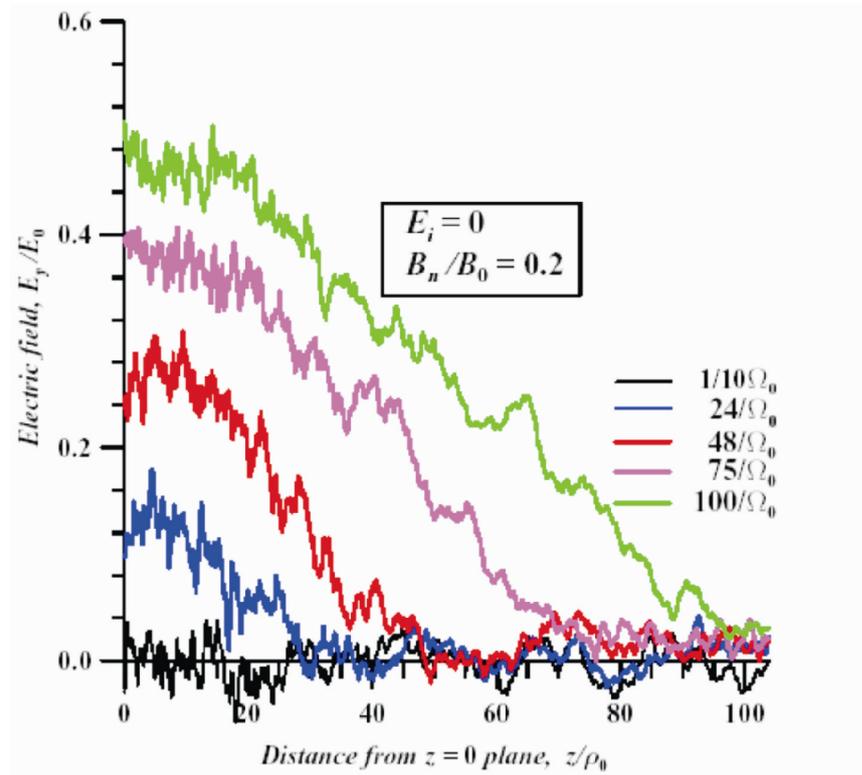

Fig. 3.4. Electric field at the early time moments, for $B_n / B_0 = 0.20$.

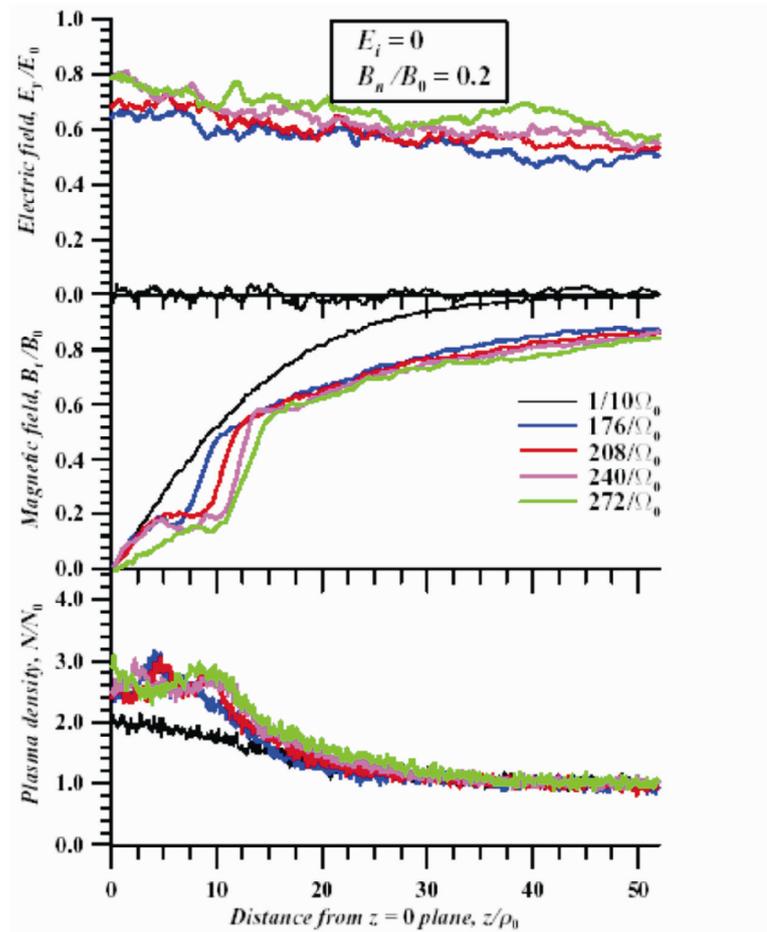

Fig. 3.5 Electric and magnetic fields and the plasma density at later time moments, for $B_n / B_0 = 0.20$.

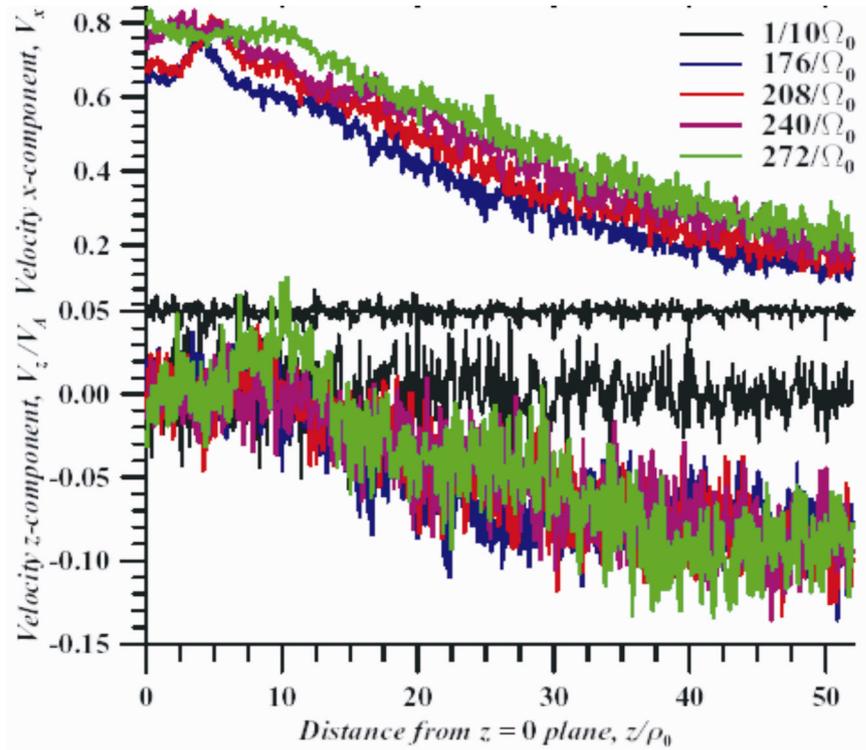

Fig. 3.6. Components of the bulk velocity at later time moments, for $B_n / B_0 = 0.20$.

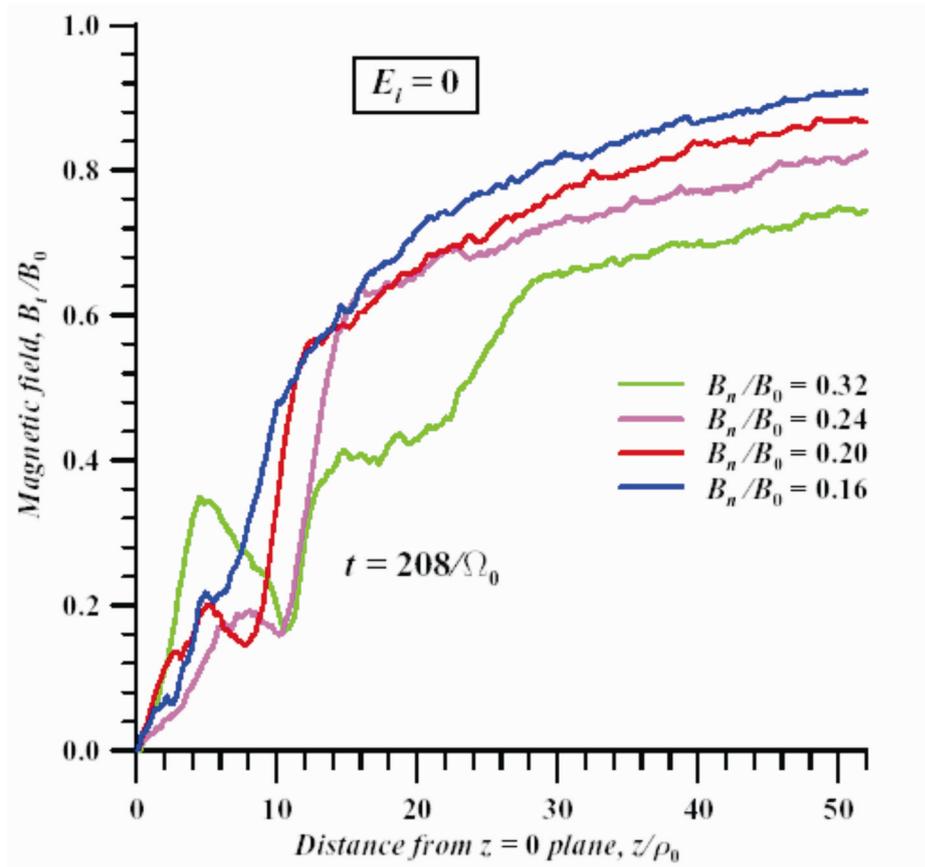

Fig. 3.7. Profiles of the magnetic field tangential component $B_t(Z)$ at the time $t = 208$ for $B_n / B_0 = 0.32,\ 0.24,\ 0.20,\ 0.16$.

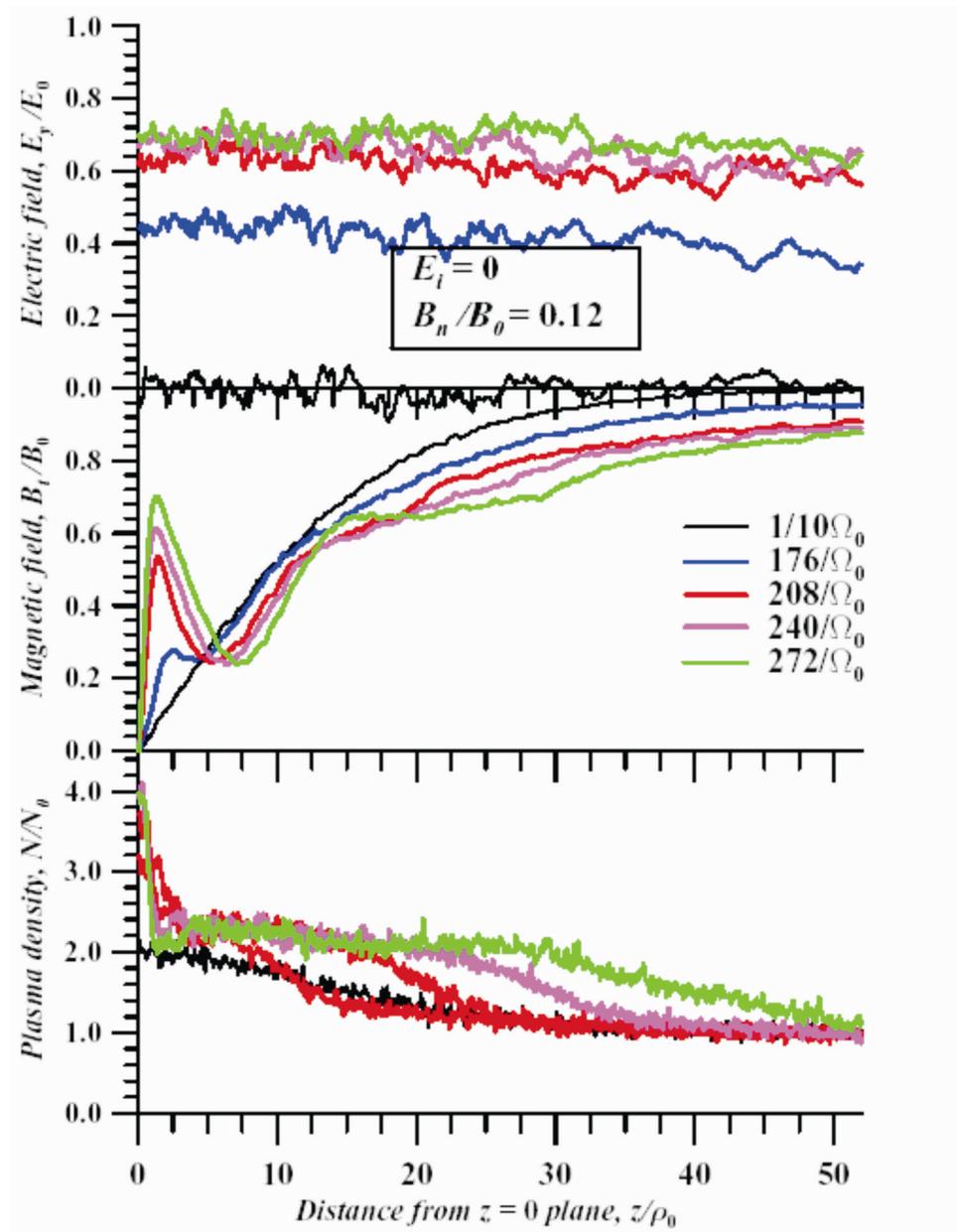

Fig. 3.8. Electric and magnetic fields and the plasma density at a number of time moments, for $B_n / B_0 = 0.12$.

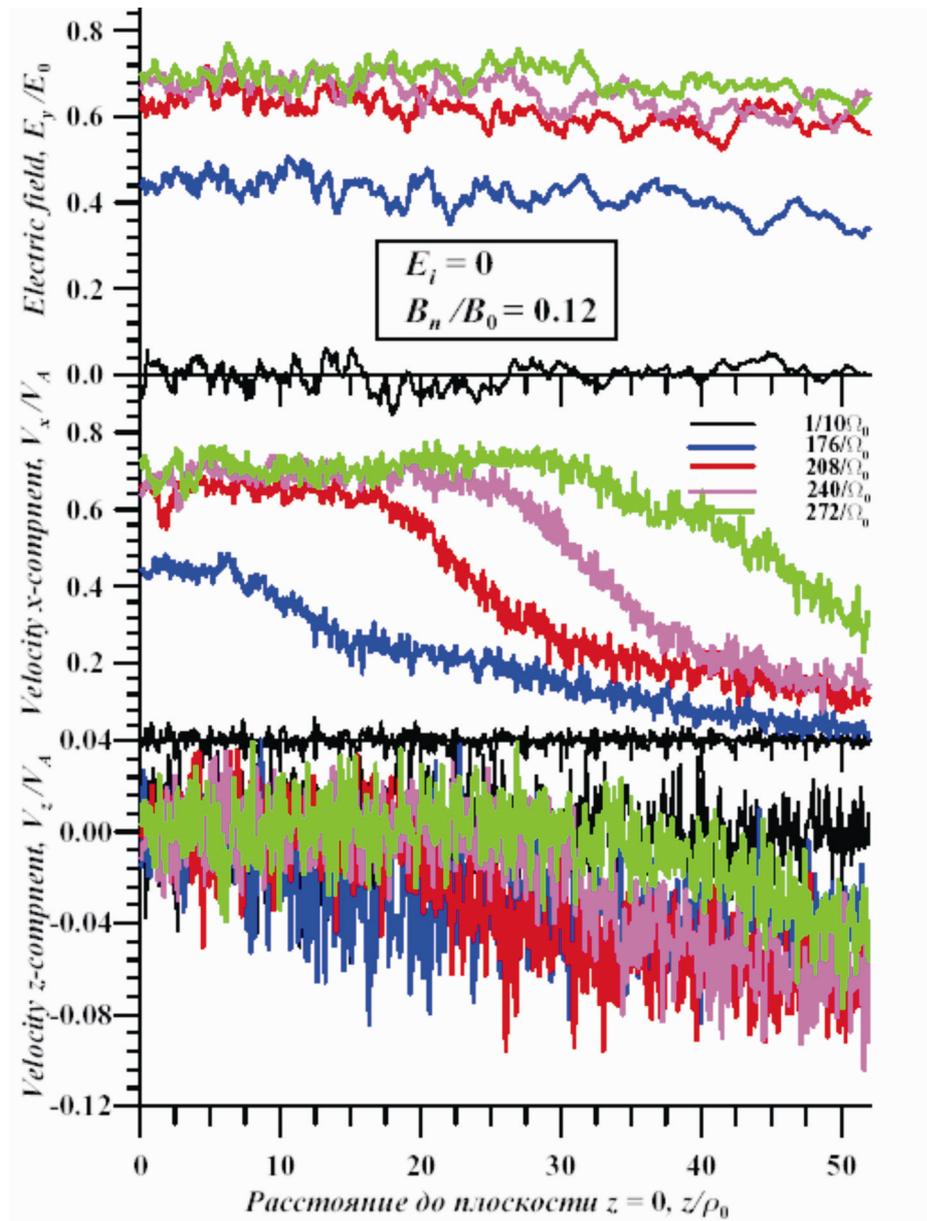

Fig. 3.9. Electric field and components of the bulk velocity at a number of time moments, for $B_n / B_0 = 0.12$.

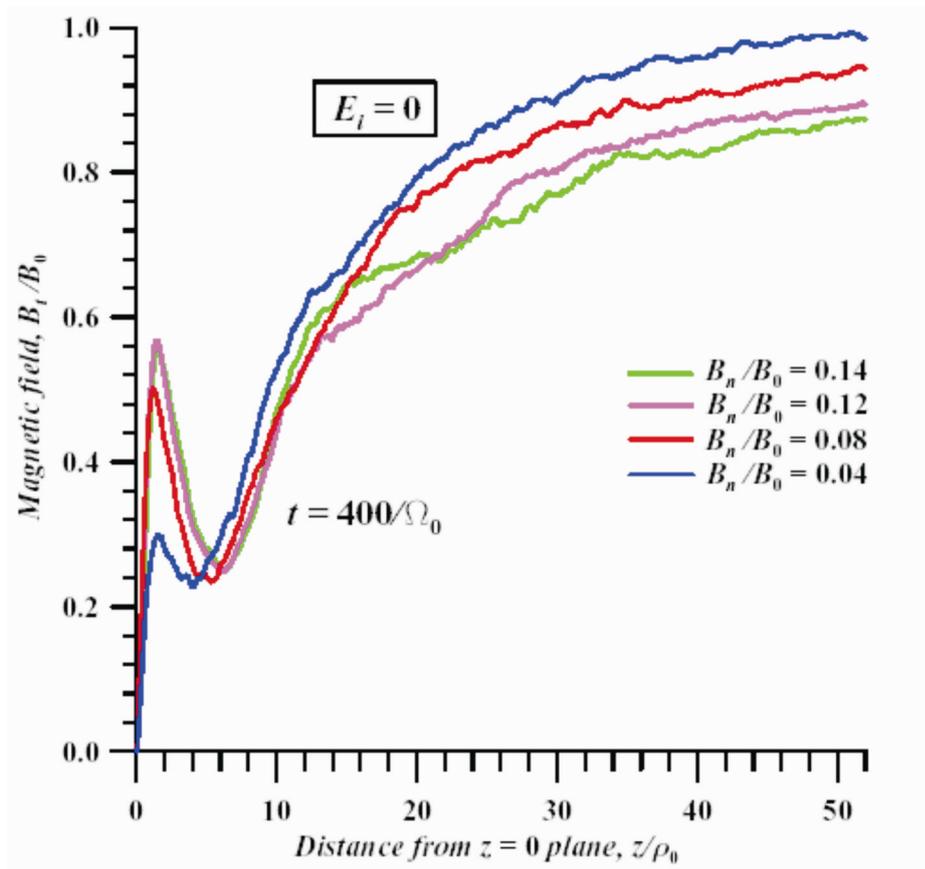

Fig. 3.10. Formation of the IKCS at various values of the magnetic field normal component lower than the critical one.

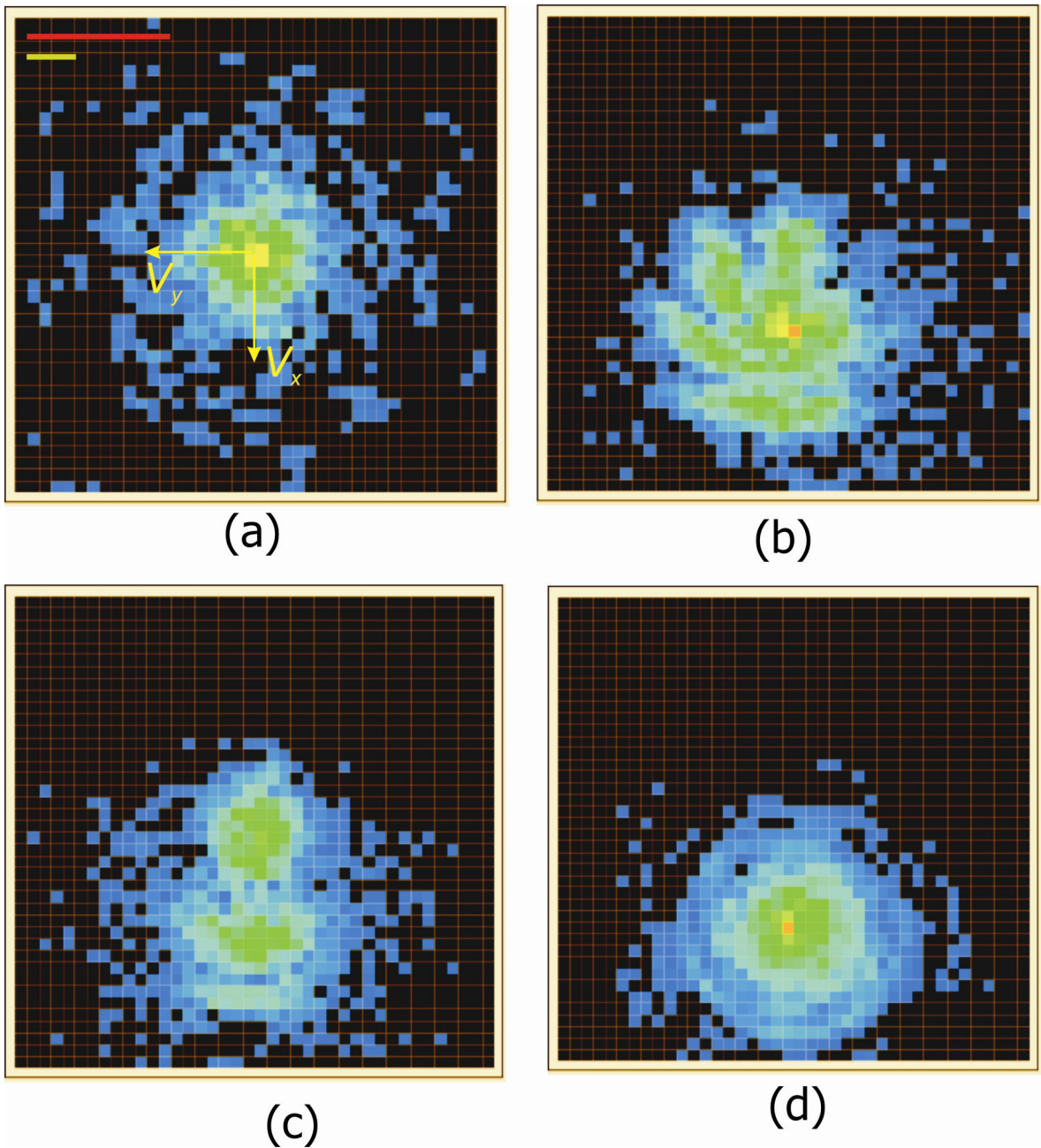

Fig. 3.11. Examples of distribution functions: phase space densities at $v_z = 0$; $v_x = v_y = 0$ at the center of each plot. (a) Initial state at $B_n / B_0 = 0.12$, $z = 0$. (b) Later moment $t = 4/\Omega_0$; $B_n / B_0 = 0.12$, $z = 0$. (c) $t = 4/\Omega_0$; $B_n / B_0 = 0.12$, $z = 12.2\rho_0$. (d) $t = 4/\Omega_0$; $B_n / B_0 = 0.2$, $z = 0$. The red bar is the value of the hot ions thermal velocity, the green one – that of the cold ions. Density decrease is denoted by the color sequence, from red to blue; black denotes negligible density.

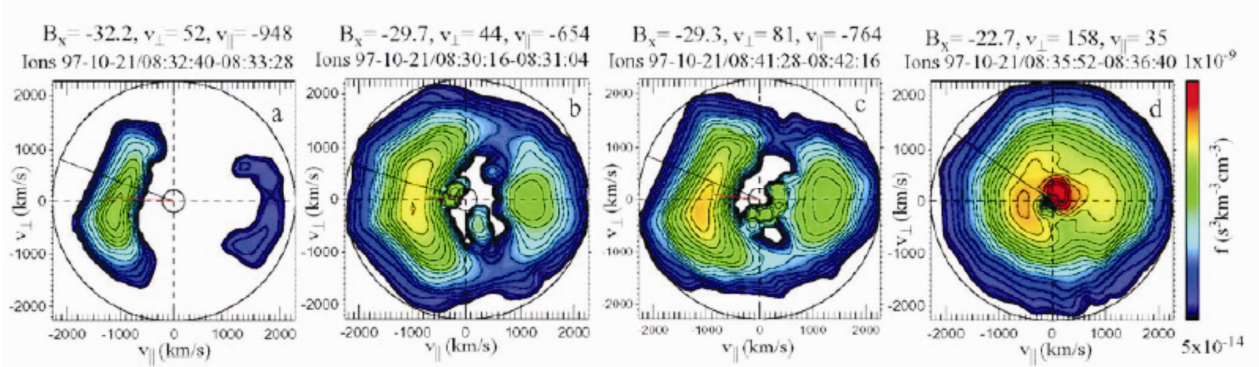

Fig. 3.12. Two-dimensional cuts of 3D ion distributions in the WIND data: anisotropy involving a field-aligned beam; $0x \parallel \mathbf{B}$; $0y \parallel [\mathbf{E} \times \mathbf{B}]$. The black line is in sunward direction. The red line indicates the bulk velocity. The distributions are taken at times ordered by decreasing $||B_x|$ (i.e., decreasing distance from the neutral sheet). (Adapted from (Raj et al 2002).)

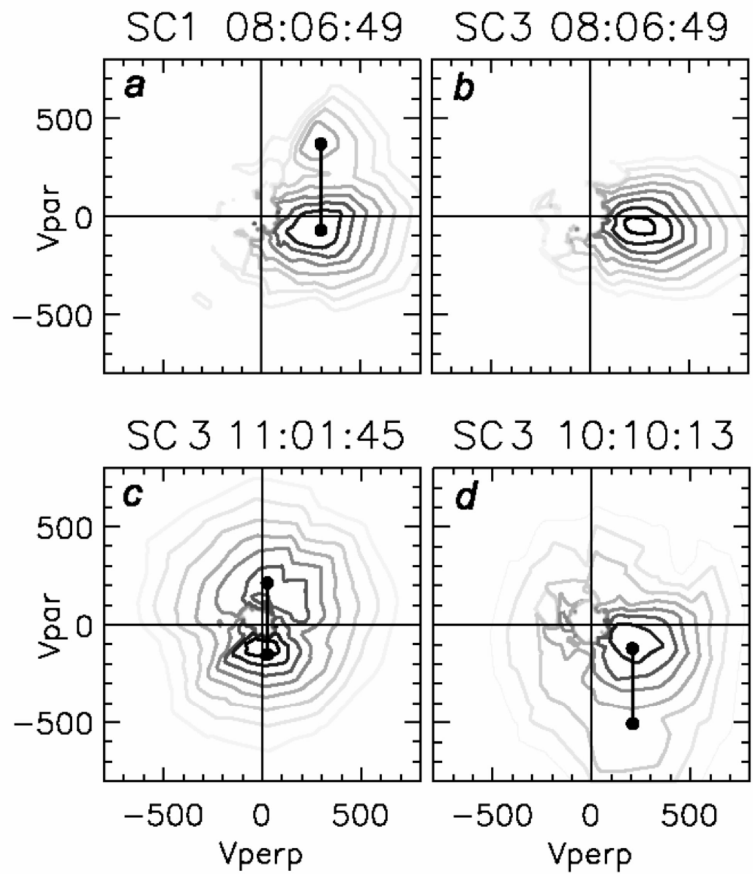

Fig. 3.13. Four cuts of the ion distribution functions in the $v_\parallel - v_\perp$ plane obtained in the CLUSTER experiment. (top) Two simultaneous measurements (a) by SC1 (in the MSBL) and (b) by SC3 (in the magnetosheath); (bottom) taken by SC3 in the MSBL (c) sunward of the reconnection site, and (d) tailward of it. Of the two populations in the MSBL, the one with the larger phase space density is the incident magnetosheath population, and the other is the reflected magnetosheath population. The vertical segments between the two populations represent the expected separation ($2V_A$).(Adapted from (Bavassano Cattaneo et al 2006).)

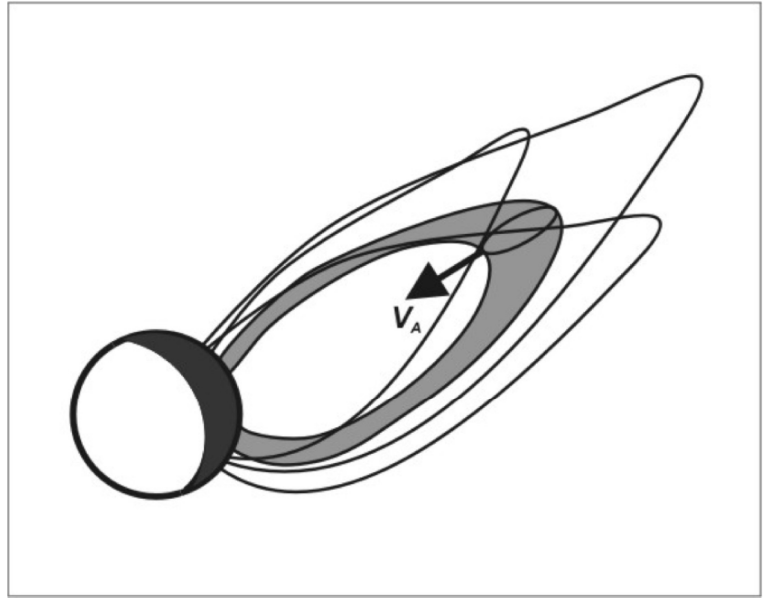

Fig. 3.14. Schematic of the "jet" motion through the field tubes being at rest in the geomagnetic tail.